\documentclass{ieeeaccess}
\usepackage{cite}
\usepackage{amsmath,amssymb,amsfonts}
\usepackage{algorithmic}
\usepackage{graphicx}
\usepackage{textcomp}
\usepackage{xcolor}
\usepackage{url}
\usepackage{bm}
\usepackage{multirow}
\usepackage[inline]{enumitem}
\usepackage{subfigure}
\usepackage{makecell}

\def\BibTeX{{\rm B\kern-.05em{\sc i\kern-.025em b}\kern-.08em
    T\kern-.1667em\lower.7ex\hbox{E}\kern-.125emX}}
\begin{document}
\history{Date of publication xxxx 00, 0000, date of current version xxxx 00, 0000.}
\doi{10.1109/ACCESS.2017.DOI}

\title{Federated Cell-Free MIMO in Non-Terrestrial Networks: Architectures and Performance}
\author{\uppercase{Alessandro Guidotti}\authorrefmark{1}, \IEEEmembership{Member, IEEE},
\uppercase{Alessandro Vanelli-Coralli\authorrefmark{2}, \IEEEmembership{Senior Member, IEEE}, and Carla Amatetti\authorrefmark{2}, \IEEEmembership{Member, IEEE}}
}
\address[1]{Consorzio Nazionale Interuniversitario per le Telecomunicazioni (CNIT), 43124 Parma, Italy, Research Unit of the University of Bologna, 40136 Bologna, Italy (e-mail: a.guidotti@unibo.it)}
\address[2]{Department of Electrical, Electronic, and Information Engineering, University of Bologna, 40136 Bologna, Italy (e-mail: \{alessandro.vanelli, carla.amatetti2\}@unibo.it)}
\tfootnote{This work has been funded by the European Commission Horizon 2020 Project DYNASAT (Dynamic Spectrum Sharing and Bandwidth-Efficient Techniques for High-Throughput MIMO Satellite Systems) under Grant Agreement 101004145. The views expressed are those of the authors and do not necessarily represent the project. The Commission is not liable for any use that may be made of any of the information contained therein.}

\markboth
{A. Guidotti \headeretal: Federated Cell-Free MIMO in Non-Terrestrial Networks: Architectures and Performance}
{A. Guidotti \headeretal: Federated Cell-Free MIMO in Non-Terrestrial Networks: Architectures and Performance}

\corresp{Corresponding author: Alessandro Guidotti (e-mail: a.guidotti@unibo.it).}

\begin{abstract}
While 5G networks are being rolled out, the definition of the potential 5G-Advanced features and the identification of disruptive technologies for 6G systems are being addressed by the scientific and academic communities to tackle the challenges that 2030 communication systems will face, such as terabit-capacity and always-on networks. In this framework, it is globally recognised that Non-Terrestrial Networks (NTN) will play a fundamental role in support to a fully connected world, in which physical, human, and digital domains will converge. In this framework, one of the main challenges that NTN have to address is the provision of the high throughput requested by the new ecosystem. In this paper, we focus on Cell-Free Multiple Input Multiple Output (CF-MIMO) algorithms for NTN. In particular: i) we discuss the architecture design supporting centralised and federated CF-MIMO in NTN, with the latter implementing distributed MIMO algorithms from multiple satellites in the same formation (swarm); ii) propose a novel location-based CF-MIMO algorithm, which does not require Channel State Information (CSI) at the transmitter; and iii) design novel normalisation approaches for federated CF-MIMO in NTN, to cope with the constraints on non-colocated radiating elements. The numerical results substantiate the good performance of the proposed algorithm, also in the presence of non-ideal information.
\end{abstract}


\titlepgskip=-15pt

\maketitle

\section{Introduction}
\label{sec:introduction}
\PARstart{D}{uring} the last years, telecommunication networks experienced an unprecedented request for an ever increasing throughput, combined with the need to support very diverse services with heterogeneous performance requirements in terms of data rate and latency. While Ultra Reliable and Low
Latency Communications, massive Machine Type Communications, and enhanced Mobile Broadband 5G services are being provided with global benefit for both economy and society, the design of new features for 5G-Advanced (5G-A) and the research on 6G technologies are already on-going, \cite{9349624, 9210567, 9040264, 9335927}. Since 2021, ITU-R initiated the development of the vision for IMT-2030 and beyond within Working Party (WP) 5D, \cite{itu_workshop}; these activities are being performed in synergy with the ITU-T Focus Group Technologies for Network 2030 (FG-NET-2030), which, between 2018 and 2020, identified a preliminary set of target services for 6G communications, \cite{itut_fg_net}. The envisioned 6G system will support a fully connected world, characterised by the convergence of the physical, human, and digital domains, \cite{itu_r_1, itu_r_2}. According to the 6G Infrastructure Association (6G-IA), three broad classes of services can be foreseen, \cite{5G_IA}: i) \emph{digital twinning}, of the systems, with actuators and sensors tightly synchronising the above mentioned domains to create digital twins of cities, factories, or even bodies; ii) \emph{connected intelligence}, in which the network serves as the cornerstone through which trusted Artificial Intelligence (AI) functions can manage the virtual representations in the digital domain; and iii) \emph{immersive communications}, in which high/ultra-high resolution visual/spatial, tactile/haptic, and other sensory data can be exchanged to create a fully immersive experience. 

In the above context, service ubiquity and continuity are critical features that only the full and seamless integration of terrestrial and Non-Terrestrial Networks (NTN) can enable, \cite{9508471, 9221119, 9502642, 8473417, 8473415, 8626457, GUIDOTTI2020107588}. The NTN segment completes the overall system architecture by providing a ubiquitous, continuous, flexible, and resilient infrastructure for: i) direct connectivity to smartphones in outdoor and in light indoor/in-vehicle (emergency communications) scenarios; ii) connectivity to mobile platforms (trains, planes, ships, drones); iii) broadcast/multicast services; iv) low latency communications to support vertical markets (\emph{e.g.}, railway, automotive, aeronautical); v) network-based positioning; and vi) Internet of Things (IoT) applications. One of the key enablers of this communication infrastructure will be the support for high throughput communications. Within NTN, current Geostationary Earth Orbit (GEO) High Throughput Satellite (HTS) systems provide hundred of Gbps through multi-beam coverage, \cite{newsat, Kyrgiazos, 5781901}. These systems are based on multi-colour, \emph{e.g.}, 3 or 4 colours, frequency reuse schemes in which the available bandwidth is split into multiple non-overlapping spectrum chunks and assigned on a geographic basis to limit interference. However, further improvements are needed to achieve the envisioned Very High Throughput Satellites (VHTS) with terabit-capacity. Several commercial endeavours are targeting the deployment of GEO systems with thousands of spot-beams, \cite{viasat}. Moreover, also Low Earth Orbit (LEO) mega-constellations, which can significantly reduce the propagation delay, have been receiving increasing interest and some of them have started the services \cite{spacex, oneweb}. Since current Physical layer (PHY) technologies already achieve a spectral efficiency close to the theoretical Shannon limit, the emphasis for future NTN systems is being placed on system design approaches aiming at increasing the exploitation of the available spectrum. This can be achieved by means of advanced spectrum usage paradigms, \emph{e.g.}, Cognitive Radios, \cite{7248435, sat_1197}, or Dynamic Spectrum Access, \cite{9221119}, or by decreasing the frequency reuse factor down Full Frequency Reuse (FFR). Notably, the latter shall be combined with effective interference management techniques, such as beamforming, precoding, and Multiple Input Multiple Output (MIMO) to exploit the massive generated co-channel interference.

In the past years, the implementation of beamforming techniques in NTN has been extensively addressed, \cite{eucnc_2022, 9739195, 9815569, EURECOM_1718, 5473886, 4686703, daniel_s2x, dimitrios_2013, 6184256, CHRISTOPOULOS201583, 5198782, 6134087, 8690982, 9438169, 6843054, 6133895, 4840357, 7811843, 6870443, 7765141, 8353925, 8580877, my_clustering, 8510728, 9684941, 8385462, 8911646}. These works, as detailed in the literature review section, focused on the increase of the system throughput in different scenarios, including unicast or multicast transmissions, ideal and non-ideal Channel State Information (CSI) at the transmitter, Geosynchronous (GSO) and, more recently, Non-GeoSynchronous Orbit (NGSO) systems, and advanced Radio Resource Management (RRM) algorithms. In this paper, we advance from that by addressing the design and performance of Cell-Free (CF) MIMO in Non-Geosynchronous Orbit (NGSO) based NTN, considering both centralised architectures, \emph{i.e.}, MIMO with co-located radiating elements on-board a single satellite, and federated solutions, \emph{i.e.}, MIMO with non co-located radiating elements on-board multiple satellites belonging to the same formation. 

\subsection{Literature review}
The literature on the application of MIMO supported by precoding and digital beamforming to NTN systems is extensive; in fact, based on the impressive benefit brought by MIMO solutions to terrestrial communications, their application to Satellite Communications has been one of the most discussed research areas in the past years. Initially, the considered Multi-User MIMO (MU-MIMO) techniques were based on the implementation of Zero Forcing (ZF) and Minimum Mean Square Error (MMSE) MIMO in satellite scenarios, \cite{EURECOM_1718}; this work showed that throughput gains in the order of $80\%$ could be obtained on both the forward and the return links. In \cite{5473886}, the authors provide a detailed and complete survey on the application of MIMO techniques over satellite channels; both fixed and mobile satellite communications were addressed, also identifying the most impacting channel impairments. In \cite{4686703}, the authors discuss the availability of only partial, and not full, CSI at the transmitter side, which is one of the most critical challenges in satellite-based MIMO. In addition to this valuable insight, the authors also introduce a novel MIMO scheme aimed at increasing the sum-rate and availability. Building on this momentum, also several projects funded by the European Space Agency (ESA) addressed the implementation of precoding to the DVB-S2X standard, \cite{daniel_s2x}; more specifically, the following practical challenges arising for MIMO in HTS systems were discussed: framing issues, non-ideal phase estimates, non-ideal CSI at the transmitter due to imperfect estimation at the user terminal, and the impact of multiple gateways. The practical impairments in the application of MIMO to DVB-S2X based systems were also discussed in \cite{dimitrios_2013} and the studies in \cite{6184256, CHRISTOPOULOS201583} provides a thorough review of precoding solutions for multi-beam satellite systems; in these latter works, the optimisation of the precoder design with linear and non-linear power constraint is also discussed. In \cite{5198782}, the non-linear Tomlinson-Harashima precoding is proposed. The performance of linear precoding, when also taking into account the traffic demand, is discussed in \cite{6134087}, where generic linear constraints were included in the transmit covariance matrix, yielding to gains compared to traditional multi-colour frequency reuse schemes as large as $170$\%. On-board Beamforming (OBBF) solutions for MIMO were discussed in \cite{8690982} for multiple gateway systems, also proposing potential solutions to mitigate the inter-beam and inter-forward link interference. In \cite{9438169}, some of the authors of this paper provided a detailed system design trade-off analysis for MMSE precoding with adaptive antennas in terms of average spectral efficiency and outage probability. In \cite{9739195}, the authors proposed a traffic-driven beam design combined with user scheduling for precoding in GEO systems. The application of unsupervised Machine Learning (ML) techniques for scheduling in precoded GEO systems is discussed in \cite{9815569}.

 More recently, multicast precoding has also been addressed. Initial studies were mainly oriented towards regularised channel inversions in which the users are served as a single terminal with an equivalent channel matrix equal to the average of the single channel matrices, \cite{6843054}. A pragmatic approach in which the linear precoding and ground-based beamforming are jointly optimised and computed at the ground segment is discussed in \cite{6133895}. In \cite{4840357}, the precoding matrix is computed through a Singular Value Decomposition (SVD). A preliminary assessment of the challenges in optimally grouping the users in multicast precoding is provided in \cite{7811843, 6870443, 7765141, 8353925}. In \cite{8580877}, a robust multi-group multicast precoding algorithm is proposed in the presence of outdated CSI. Some of the authors of this paper proposed a thorough analysis of users grouping in multicast precoding by modelling it as a clustering problem, \cite{my_clustering}; in addition, novel clustering algorithms, both for variable and fixed cluster sizes, are proposed showing significant performance improvements. The same authors proposed a geographical scheduling for unicast and multicast precoding, based on serving together only users that belong to the same zone within the corresponding reference beam, \cite{8510728}.
 
 Finally, it is worthwhile mentioning that, recently, the application of precoding on the feeder link in the presence of multiple gateways has also been addressed, \cite{9684941, 8385462, 8911646}. 
 
 \begin{figure*}[t!]
	\centering
	\includegraphics[width=0.75\textwidth]{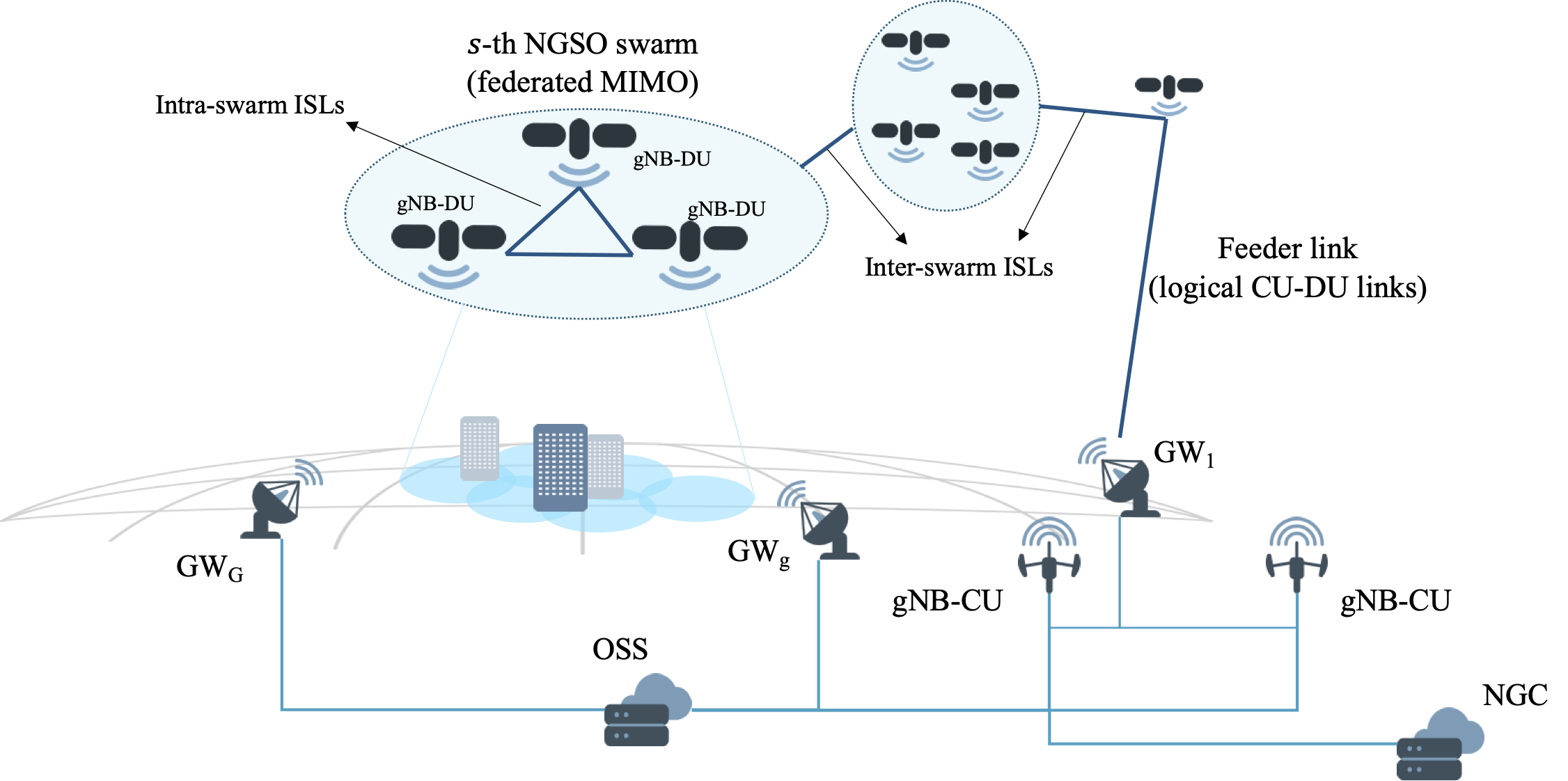}
	\caption{System architecture for federated CF-MIMO in NGSO-based NTN with regenerative payloads, functional split, and intra-/inter-swarm ISLs.}
	\label{fig:architecture}
\end{figure*}

\subsection{Paper contribution and organisation}
To the best of our knowledge, the extensive and valuable studies in the available literature focused on MIMO in NTN for CSI-based algorithms, with centralised architectures, \emph{i.e.}, all radiating elements colocated on the same satellite; moreover, architecture aspects are seldom discussed. In this work, inspired by the initial analyses that the authors reported in \cite{eucnc_2022}, and moving from the current State-of-the-Art, we
\begin{itemize}
	\item design a novel location-based CF-MIMO algorithm for NTN NGSO constellations which is completely user-centric, \emph{i.e.}, tailored to the users' and not based predetermined beam lattices, and applicable in both centralised and federated architectures;
	\item provide a unified mathematical framework for both Cell-Free (user-centric) and beam-based MIMO through federated NTN NGSO nodes;
	\item design and thoroughly discuss the architecture design choices allowing the implementation of federated Cell-Free and beam-based MIMO solutions in NTN NGSO constellations. Both regenerative, with functional split options, and transparent payloads are considered;
	\item design novel power normalisation approaches for federated MIMO algorithms that can be applied to swarms of NGSO nodes;
	\item assess the performance taking into account different sources of non-ideal knowledge at the transmitter for the computation of the beamforming matrices, including non ideal location estimation and modelling errors of the radiation pattern.
\end{itemize}
The remainder of this paper is as follows: i) in Section~\ref{sec:architecture}, we discuss the different architecture options and define the considered system architecture; ii) in Section~\ref{sec:model_beam} we describe the system model and the main assumptions; iii) in Section~\ref{sec:assessment}, the numerical assessment is provided with both ideal and non-ideal knowledge of the required information at the transmitter side; finally, Section~\ref{sec:conclusions} concludes this work.

\subsection{Notation}
Throughout this paper, and if not otherwise specified, the following notation is used: bold face lower case and bold face upper case characters denote vectors and matrices, respectively. $\mathbf{a}_{i,:}$ and $\mathbf{a}_{:,i}$ denote the $i$-th row and the $i$-th column of matrix $\mathbf{A}$, respectively. ${\left(\cdot\right)}^{-1}$ denotes the matrix inversion operator. ${\left(\cdot\right)}^T$ denotes the matrix transposition operator. ${\left(\cdot\right)}^H$ denotes the matrix conjugate transposition operator. $\mathrm{diag}\left(\mathbf{a}\right)$ denotes a diagonal matrix with the vector $\mathbf{a}$ on its main diagonal. $\mathbf{I}_K$ denotes the identity matrix of order $K$. $\mathrm{tr}\left(\mathrm{A}\right)$ denotes the trace of matrix $\mathrm{A}$.

\begin{table*}[t!]
\renewcommand{\arraystretch}{1.3}
\centering
\caption{Architecture options for MIMO in NGSO NTN.}
\label{tab:architecture}
 \begin{tabular}{|c|c|c|c|c|c|c|} 
 \hline
 \textbf{Payload type} & \multicolumn{3}{c|}{\textbf{Architecture option}} & \textbf{Computation} & \textbf{Application} & $\Delta t$ \textbf{factors} \\
 \hline
 \multirow{2}{*}{Transparent} & \multirow{2}{*}{OGC} & OGBF & \multirow{2}{*}{centralised} & \multirow{2}{*}{gNB (on-ground)} & gNB (on-ground) & \multirow{2}{*}{user+feeder(+ISLs)} \\ \cline{3-3}\cline{6-6}
  & & OBBF & &  & on-board & \\
\hline
\multirow{3}{*}{Regenerative} & \multirow{2}{*}{OGC} & OGBF & \multirow{3}{*}{\makecell{federated or\\ centralised}} & \multirow{2}{*}{gNB-CU (on-ground)} & gNB-CU (on-ground) & \multirow{2}{*}{user+feeder(+ISLs)} \\   \cline{3-3}\cline{6-6}
  & & OBBF & &  & gNB-DU (on-board) & \\ \cline{2-3} \cline{5-7}
  & OBC & OBBF & & gNB-DU (on-board) & gNB-DU (on-board) & user(+ISLs)$^1$\\   \hline  
\multicolumn{6}{l}{$^1$: The ISLs with OBC solutions are intra-swarm and, thus, might be negligible in terms of additional latency.}
 \end{tabular}
\end{table*}

\section{System Architecture}
\label{sec:architecture}
In this Section, we provide a thorough description on the NTN architecture design  to support MIMO solutions. Notably, the implementation of beamforming algorithms, detailed in Section~\ref{sec:model}, is based on the knowledge of CSI or location information at the transmitter side that shall be provided by the user terminals. Different architecture options can be considered depending on where the beamforming coefficients are computed and where they are applied to the users' signals, which is defined also based on the payload capabilities. Section~\ref{sec:architecture_1} describes the system architecture and the available design options, while in Section~\ref{sec:architecture_2} we discuss how they impact the beamforming algorithms due to the different signalling latencies.
\subsection{Architecture design options}
\label{sec:architecture_1}
To support CF-MIMO in NTN, the satellite system architecture is impacted by many design choices, such as: i) the type of satellite payload, \emph{i.e.}, regenerative or transparent; ii) the type of functional split when regenerative payloads are assumed, \emph{i.e.}, which layers of the NR gNB are implemented on-board in the Distributed Unit (gNB-DU) and which ones are implemented on-ground in the Centralised Unit (gNB-CU); iii) the network entity in which the beamforming coefficients are computed based on the considered CF-MIMO algorithm; and iv) the network entity in which the beamforming coefficients are applied to the signals, \emph{i.e.}, On-Board Beamforming (OBBF) or On-Ground Beamforming (OGBF). The system architecture is represented in Figure~\ref{fig:architecture} and it includes:
\begin{itemize}
	\item The ground segment, which includes $G$ on-ground gateways (GWs) providing NTN access to the Terrestrial Network(s) (TN). In particular, the $G$ GWs provide the connectivity between the NGSO nodes in the constellation, the gNBs, and the Next Generation Core network (NGC). As for the latter, the ground segment also includes the Operations Support Systems (OSS), which is in charge of managing the overall satellite system. Depending on the type of payload on-board the NTN nodes, different elements are needed in this segment: i) with transparent payloads, the full gNB shall be implemented on-ground and the NTN nodes basically act as relays; ii) with regenerative payloads and functional split, as shown in Figure~\ref{fig:architecture}, the gNB-DUs can be located on-board, leaving the gNB-CUs on-ground. In the latter case, it shall be mentioned that each gNB-CU (full gNB with transparent payloads) can manage up to tens of connections; assuming one connection per beam, depending on the total number of beams per NTN node, multiple gNB-CUs (full gNBs) might be needed to manage all of the connections supported by the gNB-DUs (single node).
	\item The Non-Terrestrial (NT) access segment, which includes the NGSO nodes in the constellation. We refer to nodes since the elements in the NGSO constellation can be any type of platform on one or more NGSO orbits, \emph{e.g.}, LEO satellites organised in one or more sub-constellations at different altitudes, or a High Altitude Platform Stations (HAPS) formation. As mentioned above, these nodes can implement either a transparent or a regenerative payload, depending on the cost and complexity of the target system. With respect to coverage, in 3GPP beam-based coverage solutions we might have: i) Earth-fixed beams, \emph{i.e.}, through digitally steering of the signals, the coverage area generated by each node is fixed on-ground independently of its position on the orbit (as long as it falls in the node field of view); or ii) Earth-moving beams, \emph{i.e.}, the coverage area of each node is always centered around its Sub Node Point (SNP) and, thus, the beams move on-ground along with the node on its orbit. When Cell-Free approaches are considered, the very concept of beams is not necessary anymore, as extensively discussed in Section~\ref{sec:model}.
	\item The on-ground user segment, composed by a potentially massive number of User Equipments (UE), either fixed or moving. These directly connect to the serving node(s) by means of the Uu air interface through the user access link. In this work, we consider both handheld terminals and Very Small Aperture Terminals (VSAT).
\end{itemize}

Based on the above observations, the selected functional split option, \cite{38801}, has an impact on where the users' scheduling and beamforming coefficients are computed; in particular, two architecture design options are possible: i) \emph{On-Ground beamforming Computation} (OGC), where the scheduling and coefficients computation is performed at the on-ground gNB-CU; or ii) \emph{On-Board beamforming Computation} (OBC), where these operations are performed at the on-board gNB-CUs. Moreover, with regenerative payloads, NGSO-based systems allow the implementation of \emph{federated} (distributed) MIMO solutions, in which all of the satellites in the same swarm (formation) cooperate to implement MIMO transmissions.

\subsection{Centralised and Federated MIMO}
\label{sec:architecture_2}
Table~\ref{tab:architecture} summarises the architecture options based on the payload type, where the users' scheduling and beamforming coefficients are computed (OGC or OBC), where the beamforming coefficients are applied to the users' signals (OGBF or OBBF), and whether a federated MIMO solution is possible or not. Depending on the selected option, the entity performing the different operations can be identified.

With legacy transparent payloads, scheduling and beamforming are entirely defined on-ground (OGC); then, the beamforming coefficients can be applied to the users' signals either on-board (OBBF) or on-ground (OGBF). In this scenario, no federated solution is possible and each satellite in the constellation operates as a standalone node; in fact, federated MIMO architectures require a tight time and frequency synchronisation among the cooperating satellites in the swarm, which can only be achieved by means of intra-swarm Inter-Satellite Links (ISL) with regenerative payloads.

When considering future regenerative payloads, legacy centralised architectures are clearly still possible. However, the exploitation of the advanced on-board computational capabilities supports two additional MIMO architecture options: i) OBC, in which the users' scheduling and beamforming coefficients are computed on-board; and ii) federated MIMO, in which multiple satellites can tightly synchronise to realise a distributed MIMO system. OBC solutions allow to perform all operations at the on-board gNB-DU: i) computation of the users' scheduling and beamforming coefficients; and ii) application of the beamforming coefficients to the users' signals. Such advanced capabilities allow to implement either a centralised or a federated MIMO algorithm. In the former case, each satellite in the NGSO constellation can operate as a standalone NTN node, thus leading to a centralised architecture; all of the UEs in the satellite's service area send the required ancillary information for the considered MIMO algorithm (CSI or location, as discussed in Section~\ref{sec:model}) on the return link and the satellite implements a centralised MIMO algorithm. With federated MIMO, the synchronisation and coordination among multiple NTN nodes in a single swarm is possible. In particular: i) a master gNB-DU collects the ancillary information (CSI or location) from the UEs, either directly or with the support of intra-swarm ISLs; ii) the master gNB-DU computes the users' scheduling and beamforming coefficients and sends the beamformed signals to the other satellites in the swarm; and iii) thanks to the intra-swarm ISLs, the satellites in the swarm can tightly synchronise the transmission of the beamformed signals in the time and frequency domains, leading federated MIMO via a distributed antenna system. It is worthwhile highlighting that federated MIMO is also possible with the on-ground computation of the users' scheduling and beamforming coefficients provided by OGC architectures; in this case, the NGSO satellites in the swarm still realise a tightly synchronised distributed antenna system, with the only difference being that the ancillary information from the UEs is sent to the on-ground gNB-CU, which is in charge of all computations. Then, the beamforming coefficients can be applied either on-ground (OGBF) or on-board (OBBF). 

On the one hand, the implementation of federated MIMO is challenging in terms of increased system complexity, due to the need for regenerative payloads with advanced on-board processors and tight intra-swarm synchronisation via ISLs. On the other hand, the deployment of a flying distributed antenna system allows to tackle the detrimental impact of harsh propagation environments, thanks to the spatial diversity at the transmitter.

\subsection{Ancillary information aging: OGC and OBC}
\label{sec:architecture_3}
\begin{figure}[t!]
	\centering
	\includegraphics[width=\columnwidth]{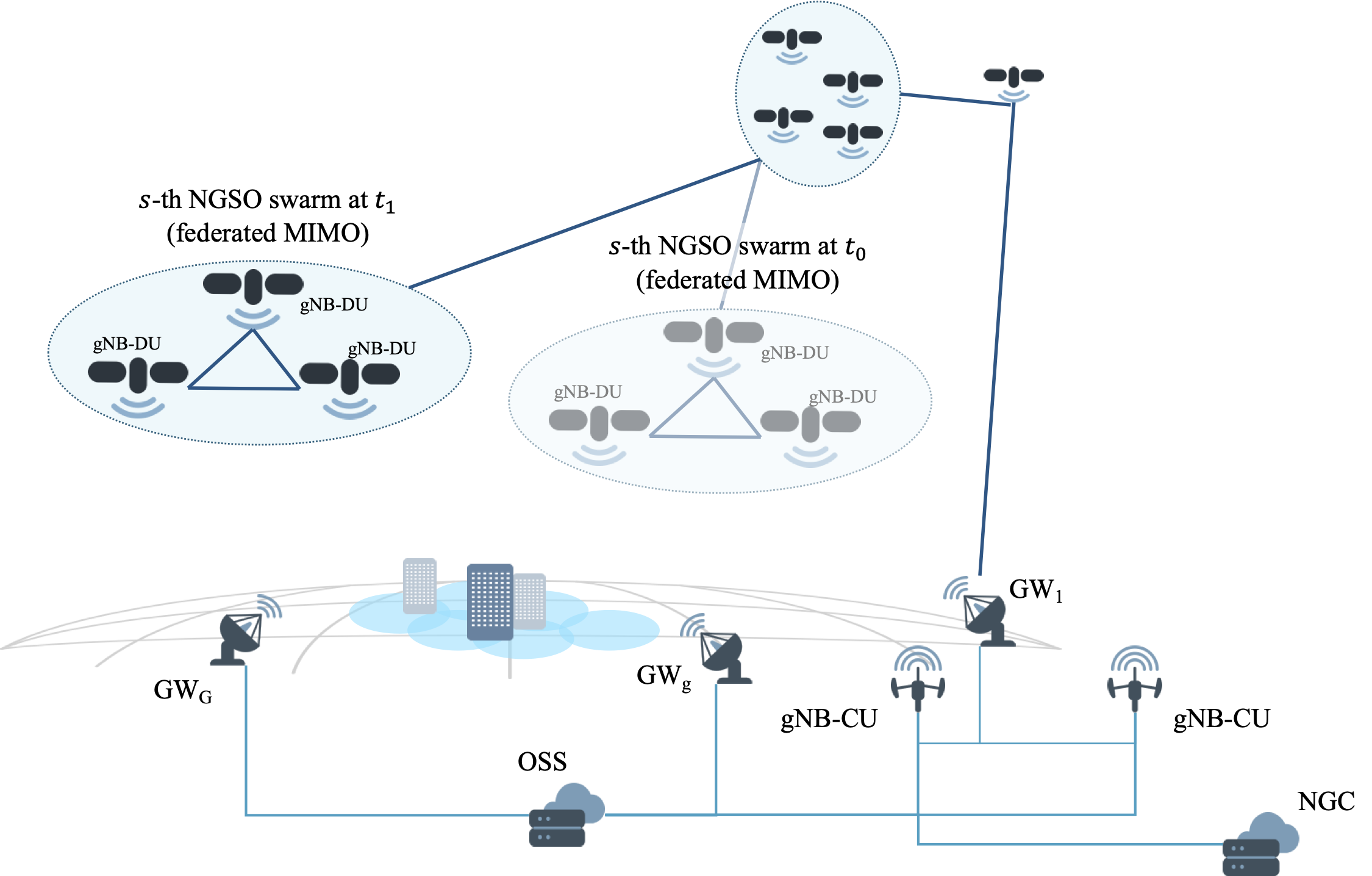}
	\caption{System architecture for federated CF-MIMO in NGSO-based NTN: estimation and transmission phases.}
	\label{fig:architecture_2}
\end{figure}

The choice between OGC and OBC is fundamental for MIMO in NTN. In fact, the MIMO algorithms considered in this work (described in Section~\ref{sec:model}) require either the CSI vectors or the location estimated at the UEs' locations to build the beamforming matrix, denoted as \emph{ancillary information}. The ancillary information is obtained by the UEs at an \emph{estimation time instant} $t_0$ and then sent to the network element computing the coefficients: the on-ground gNB(-CU) with OGC or the master on-board gNB-CU with OBC. The transmission of the beamformed signals from either the federated gNB-CUs or the standalone gNB-CU (centralised architectures) then happens at a \emph{transmission time instant} $t_1>t_0$. During the \emph{aging interval} $\Delta t = t_1-t_0$, both the NGSO nodes and the UEs have moved and, thus, there is a misalignment between the actual channel encountered during the transmission and the ancillary information used to compute the beamforming matrix. This is represented in Figure~\ref{fig:architecture_2}, where only the swarm movement is depicted for the sake of clarity. Notably, the MIMO performance is deeply impacted by any misalignment between the actual channel and the beamforming matrix; thus, the smaller the aging interval, the better the performance of federated MIMO.

When considering OGC, the users' scheduling and coefficients are computed on-ground and, thus, the aging interval can be computed as:
\begin{equation}
	\Delta t_{OGC} = \tau_{user} + \tau_{feeder}^{(DL)} + \tau_{feeder}^{(UL)} + \tau_{p} + \tau_{rout} + \tau_{ad}
\end{equation}
where: i) $\tau_{user}$ is the latency on the user return link; ii) $\tau_{feeder}^{(DL)}$ is the latency on the feeder downlink; iii) $\tau_{feeder}^{(UL)}$ is the latency on the feeder uplink; iv) $\tau_{p}$ is the processing delay to compute the users' scheduling and beamforming coefficients; v) $\tau_{rout}$ is the latency due to routing on the ISLs, if present; and v) $\tau_{ad}$ includes any additional source of latency. When OBC is considered, the aging interval is significantly reduced; in fact, all computations are performed on-board and thus:
\begin{equation}
	\Delta t_{OBC} = \tau_{user} + \tau_{p} + \tau_{ad}
\end{equation}
Compared to $\Delta t_{OGC}$, $\Delta t_{OBC}$ only includes the over-the-air latency on the user access link, in addition to the processing and additional delays. Clearly, this advantage is lost if the functional split option does not allow to perform the required computations on-board, \emph{i.e.}, regenerative payloads with a OGC approach. The factors impacting the aging interval $\Delta t$ are summarised in Table~\ref{tab:architecture}.

 It is worthwhile highlighting that the advantage of OBC is not only related to the reduction of the aging interval and, thus, improving the MIMO performance; in fact, the signalling overhead on the feeder link, and on any inter-swarm ISL that might be needed to connect the swarm with the serving gNB-CU, is massively reduced. The reduction in signalling also includes all information that is needed to implement the desired Radio Resource Management (RRM) algorithm, which might include the UEs' capacity request and type of traffic, as well as the terminal class. With respect to the latter, it shall be mentioned that this information might be classified by the manufacturers; in this case, an estimate can be identified based on ancillary terminal parameters/information. RRM aspects are not addressed in this work, without impacting the generality of the proposed architectures or algorithms. Finally, it is worthwhile highlighting that both the above advantages are achieved with both centralised and federated MIMO architectures.

\section{System Model}
\label{sec:model}
We consider a constellation with $M$ NGSO nodes providing connectivity to the on-ground UEs in the service area. Notably, for a generic coverage area, only a subset of nodes will be visible from all of the UEs, based on the nodes' field of view and on minimum elevation angle requirements. In the following, we assume that a single swarm of $N_{node}$ nodes is visible from all of the $N_{ue}$ UEs in the considered area. Each node is equipped with a Uniform Planar Array (UPA) with $N_F$ radiating elements; for the sake of simplicity, and without affecting the generality of our work, we assume that all of the nodes are at the same altitude and equipped with the same antenna configuration (\emph{i.e.}, number of radiating elements, UPA configuration, element radiation pattern).

\begin{figure}[t!]
	\centering
	\includegraphics[width=\columnwidth]{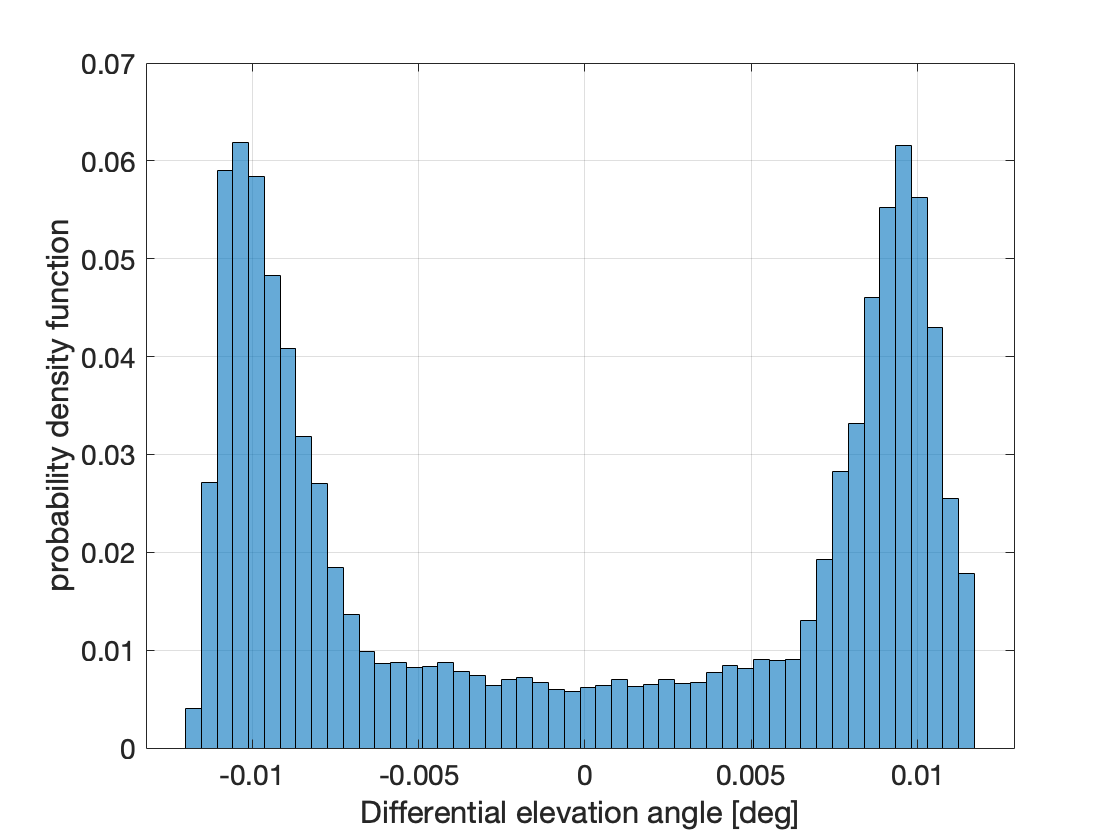}
	\caption{Probability density function (pdf) of the difference between the elevation angle at $t_1$ and the elevation angle at $t_1$, $N_{node}=2$.}
	\label{fig:diff_elev}
\end{figure}

As discussed in Section~\ref{sec:architecture_3}, the users estimate the ancillary information required for the considered MIMO algorithm (CSI or location) at the estimation time instant $t_0$; then, the actual transmission of the beamformed signals occurs at the transmission time instant $t_1$. During the aging interval, there are several sources of misalignment between the actual channel at $t_1$ and the beamforming matrix based on information obtained at $t_0$: i) the nodes moved along their orbits; ii) the UEs might have moved, depending on the terminal type; and iii) different realisations of the stochastic terms in the channel coefficients (\emph{e.g.}, large scale losses, scintillation) are present. As for the latter, the generic coefficient between the $i$-th on-ground UE and the $n$-th radiating element of the UPA on-board the $s$-th node at the generic time instant $t$ can be computed as:
\begin{equation}
\label{eq:channel_coeff}
	h_{i,n,s}^{(t)} = \frac{g_{i,n,s}^{(TX,t)}g_{i,n,s}^{(RX,t)}}{4\pi\frac{d_{i,s}^{(t)}}{\lambda}\sqrt{L_{i,s}^{(t)}\kappa B T_i}}e^{-\jmath\frac{2\pi}{\lambda}d_{i,s}^{(t)}}e^{-\jmath\varphi_{i,s}^{(t)}}
\end{equation}
where: i) $d_{i,s}^{(t)}$ is the slant range between the $i$-th user and the $s$-th node, which is assumed to be the same for all the co-located radiating elements on-board the node; ii) $\lambda$ is the signal wavelength; iii) $\kappa B T_i$ denotes the thermal noise power, with $B$ being the user bandwidth (for simplicity assumed to be the same for all users) and $T_i$ the equivalent noise temperature of the $i$-th receiver; iv) $L_{i,s}^{(t)}$ represents the additional losses between the $s$-th node and the $i$-th user, assumed to be the same for all the co-located radiating elements on-board the node; v) $g_{i,n,s}^{(TX,t)}$ and $g_{i,n,s}^{(RX,t)}$ represent the transmitting and receiving complex antenna patterns between the $i$-th user and the $n$-th radiating element on-board the $s$-th node, respectively; and vi) $\varphi_{i,s}^{(t)}$ is the phase misalignment that might be present between different nodes due to non-ideal swarm synchronisation, modelled as a Gaussian random variable (r.v.) $\mathcal{N}\left(0,2\pi\right)$. The additional losses are computed based on TR 38.811, \cite{38811}:
\begin{equation}
	L_{i,s}^{(t)} = L_{i,s}^{(SHA,t)} + L_{i,s}^{(ATM,t)} + L_{i,s}^{(SCI,t)} + L_{i,s}^{(CL,t)}
\end{equation}
in which: i) $L_{i,s}^{(SHA,t)}$ denotes the log-normal shadowing loss with standard deviation $\sigma_{SHA}$; ii) $L_{i,s}^{(ATM,t)}$ includes the atmospheric loss due to gaseous absorption; iii) $L_{i,s}^{(SCI,t)}$ is the scintillation loss; and iv) $L_{i,s}^{(CL,t)}$ is the Clutter Loss (CL), to be included for UEs in Non-Line-of-Sight (NLOS) conditions. Referring to the 3GPP channel model, the UE is defined to be in LOS or NLOS conditions with a probability that is a function of the elevation angle and the propagation environment (sub-urban, urban, dense-urban). In this context, we assume that a UE that is LOS (NLOS) conditions during the estimation phase is still in LOS (NLOS) conditions in the transmission phase. This assumption is motivated by observing that the probability that the propagation conditions of the UE will change from LOS (NLOS) to NLOS (LOS) after a few ms is negligible; in fact, as shown in Figure~\ref{fig:diff_elev}, the values of the differential elevation angle between $t_0$ and $t_1$ are negligible\footnote{These results where obtained with $N_{node}=2$ and the system configuration described in Section~\ref{sec:assessment}.}, considering that the probabilities of LOS or NLOS conditions are provided with a $10^{\circ}$ granularity in \cite{38811}. This assumption implies that the UE has the same CL and $\sigma_{SHA}$ in both the estimation and transmission phases, but the realisations of the log-normal r.v. modelling the shadowing are different.

In the following, the estimated CSI vector between the $s$-th node and the $i$-th user is represented by $\mathbf{h}_{i,s}^{(t_0)} = \left[h_{i,1,s}^{(t_0)},\ldots,h_{i,N_F,s}^{(t_0)}\right]$. For the generic $i$-th user, its overall channel signature can be obtained by collecting the CSI vectors from all of the NGSO nodes into the $N_{node}N_{F}$-dimensional $\mathbf{h}_{i,:}^{(t_0)} = \left[\mathbf{h}_{i,1}^{(t_0)}, \ldots, \mathbf{h}_{i,N_{node}}^{(t_0)}\right]$. Finally, the overall $N_{ue}\times\left(N_{node}N_{F}\right)$  channel matrix at the estimation time $t_0$ is given by $\mathbf{H}^{(t_0)}_{sys} = {\left[{\left(\mathbf{h}_{1,:}^{(t_0)}\right)}^T,\ldots,{\left(\mathbf{h}_{N_{node},:}^{(t_0)}\right)}^T\right]}^T$. For each time slot, the RRM scheduling function $\mathcal{S}$ provides a subset of $K$ users to be served, leading to a $K\times\left(N_{node}N_{F}\right)$ channel matrix $\mathbf{H}^{(t_0)} = \mathcal{S}\left(\mathbf{H}_{sys}^{(t_0)}\right)\subseteq\mathbf{H}^{(t_0)}_{sys}$. Based on the channel matrix estimated at $t_0$, the beamforming algorithm (detailed in the next sections) provides a $\left(N_{node}N_{F}\right)\times K$ complex beamforming matrix $\mathbf{W}^{(t_0)}=\mathcal{B}\left(\mathbf{H}^{(t_0)}\right)$, which projects the $K$-dimensional column vector $\mathbf{s} = {\left[s_1, \ldots, s_{N_{ue}}\right]}^T$ containing the unit-variance user symbols into the $\left(N_{node}N_{F}\right)$-dimensional space defined by all of the swarm radiating elements. The signal received by the generic $i$-th UE in FFR is given by:
\begin{equation}
\label{eq:rx_signal_1}
	y_i = \underbrace{\mathbf{h}_{i,:}^{(t_1)}\mathbf{w}_{:,i}^{(t_0)}s_i}_{\mathrm{intended}} + \underbrace{\sum_{\substack{\ell = 1\\ \ell\neq i}}^{K} \mathbf{h}_{i,:}^{(t_1)}\mathbf{w}_{:,\ell}^{(t_0)}s_{\ell}}_{\mathrm{interfering}} + z_i
\end{equation}
where $z_i$ is a circularly symmetric Gaussian r.v. with zero mean and unit variance, which is licit observing that the channel coefficients in (\ref{eq:channel_coeff}) are normalised to the noise power. From (\ref{eq:rx_signal_1}), the $K$-dimensional vector of received symbols is:
\begin{equation}
	\mathbf{y} = \mathbf{H}^{(t_1)}\mathbf{W}^{(t_0)}\mathbf{s} + \mathbf{z}
\end{equation}
As previously discussed, it can be noticed that there is a misalignment between the estimated channel matrix exploited to compute $\mathbf{W}^{(t_0)}$, function of $\mathbf{H}^{(t_0)}$, and the actual channel in the transmission phase, $\mathbf{H}^{(t_1)}$.

The Signal-to-Interference-plus-Noise Ratio (SINR) of the generic $i$-th UE can be obtained as:
\begin{equation}
\label{eq:SINR}
	\gamma_i = \frac{{\left| \mathbf{h}_{i,:}^{(t_1)}\mathbf{w}_{:,i}^{(t_0)} \right|}^2}{1 + \sum_{\substack{\ell = 1\\ \ell\neq i}}^{K} {\left| \mathbf{h}_{i,:}^{(t_1)}\mathbf{w}_{:,\ell}^{(t_0)}\right|}^2}
\end{equation}
From the above SINR, the rate achieved by the $i$-th user can be evaluated either from the Shannon bound formula or from the adopted Modulation and Coding scheme (MCS). In this framework, 3GPP TR 38.803 reports that the spectral efficiency for system-level simulations can be obtained through the following truncated form of the Shannon bound, \cite{38803}:
\begin{equation}
\label{eq:shannon_trunc}
	\eta_i = \begin{cases}
		0, \ \gamma_i<\gamma_{min} \\
		\varepsilon\cdot \log_2\left(1+\gamma_i\right), \ \gamma_{min}\leq \gamma_i< \gamma_{max}\\
		\varepsilon\cdot \log_2\left(1+\gamma_{max}\right),\ \gamma_i\geq\gamma_{max}
	\end{cases}
\end{equation}
where $\gamma_{min}=-10$ dB and $\gamma_{max}=30$ dB are the minimum and maximum SINR of the MCS, respectively, and $\varepsilon$ is an attenuation factor representing the implementation loss. Since the attenuation factor is a multiplicative term outside of the Shannon formula, in the following we assume $\varepsilon=1$ since a different value only acts as a scaling factor on all of the results discussed below, \emph{i.e.}, it does not impact the relationship among the different techniques and scenarios and the general trends.

\subsection{CSI-based CF-MIMO}
\label{sec:model_csi}
CSI-based techniques require each UE to estimate the CSI vector $\mathbf{h}_{i,:}^{(t_0)}$ during the estimation phase and to report it to the network element in charge of the computation of $\mathbf{W}^{(t_0)}$ (\emph{i.e.}, gNB-DU with OGC or gNB-DU with OBC). Notably, among these, Minimum Mean Square Error (MMSE) beamforming is the best algorithm in the sense of SINR maximisation; as such, it is considered as the upper-bound performance benchmark\footnote{This is a an equivalent and more computationally efficient formulation of the MMSE beamformer, \cite{9142191}.}:
\begin{equation}
	 \mathbf{W}_{MMSE}^{(t_0)} = {\mathbf{H}}^H\left({\mathbf{H}}{\mathbf{H}}^H + \mathrm{diag}(\boldsymbol{\alpha})I_{K}\right)^{-1}
\end{equation}
where, for the sake of clarity, we dropped the time instant $t_0$ from the channel matrix. In the above equation, $\mathrm{diag}(\boldsymbol{\alpha})$ is a vector of $K$ regularisation factors; since the channel coefficients are normalised to the noise power, the optimal value is given by $\boldsymbol{\alpha}=N_F/P_t$, \cite{5962672}, where $P_t$ is the available power per node in the swarm.

\begin{table*}[t!]
\renewcommand{\arraystretch}{1.3}
\centering
\caption{Summary of the considered beamforming algorithms.}
\label{tab:algorithms}
 \begin{tabular}{|c|c|c|c|c|} 
 \hline
 \textbf{Algorithm} & \textbf{Channel coefficient} & \textbf{Beamforming matrix} & \textbf{Information} & \textbf{Errors} \\
 \hline
 MMSE &  $h_{i,n,s}^{(t)} = \frac{g_{i,n,s}^{(TX,t)}g_{i,n,s}^{(RX,t)}}{4\pi\frac{d_{i,s}^{(t)}}{\lambda}\sqrt{L_{i,s}^{(t)}\kappa B T_i}}e^{-\jmath\frac{2\pi}{\lambda}d_{i,s}^{(t)}}e^{-\jmath\varphi_{i,s}^{(t)}}$ & ${\mathbf{H}}^H\left({\mathbf{H}}{\mathbf{H}}^H + \mathrm{diag}(\boldsymbol{\alpha})I_{K}\right)^{-1}$ & CSI & \makecell[c]{CSI estimation \\ Low-SINR estimation \\ Air interface adj. \\ UE/node movement} \\
 \hline
 LB-MMSE & $\widetilde{h}_{i,n,s}^{(t)} = \frac{\widetilde{g}_{i,n,s}^{(TX,t)}\widetilde{g}_{i,n,s}^{(RX,t)}}{4\pi\frac{\widetilde{d}_{i,s}^{(t)}}{\lambda}\sqrt{\kappa B T_i}}e^{-\jmath\frac{2\pi}{\lambda}\widetilde{d}_{i,s}^{(t)}}$ & ${\widetilde{\mathbf{H}}}^H\left(\widetilde{\mathbf{H}}{\widetilde{\mathbf{H}}}^H + \mathrm{diag}(\boldsymbol{\alpha})I_{K}\right)^{-1}$ & Location & \makecell[c]{Location estimation \\ Radiation pattern model \\ UE/node movement}\\
 \hline
 SS-MMSE & $\widetilde{h}_{i,n,s}^{(BC,t)} = \frac{\widetilde{g}_{i,n,s}^{(BC,TX,t)}\widetilde{g}_{i,n,s}^{(BC,RX,t)}}{4\pi\frac{\widetilde{d}_{i,s}^{(BC,t)}}{\lambda}\sqrt{\kappa B T_i}}e^{-\jmath\frac{2\pi}{\lambda}\widetilde{d}_{BC,i,s}^{(t)}}$ & ${\widetilde{\mathbf{H}}}^H_{BC}\left(\widetilde{\mathbf{H}}_{BC}{\widetilde{\mathbf{H}}}^H_{BC} + \mathrm{diag}(\boldsymbol{\alpha})I_{K}\right)^{-1}$ & Location & \makecell[c]{Location estimation \\ Radiation pattern model \\ Approx. location \\ UE/node movement}\\
 \hline
MB  & $b_{n,\ell} = \frac{1}{\sqrt{N_F}}e^{-\jmath k_0\mathbf{r}_n \cdot \mathbf{c}_{\ell}}$ & ${\left[b_{n,\ell}\right]}_{\substack{n=1,\ldots,N_F\\ \ell=1,\ldots,N_B}}$ & Location & \makecell[c]{Location estimation \\ Approx. location \\ UE/node movement} \\
\hline
 \end{tabular}
\end{table*}

\subsection{Location-based CF-MIMO}
\label{sec:model_location}
Inspired by the Spatially Sampled (SS-MMSE) algorithm proposed in \cite{eucnc_2022}, we design a new location-based CF-MIMO solution. In particular, when the UEs are equipped with GNSS capabilities, they can estimate their locations and provide them to the gNB-CU (OGC) or gNB-DU (OBC). This information, combined with the knowledge of the swarm ephemeris, can be exploited to infer the channel coefficients between the UE and the NGSO nodes. In particular, all terms in (\ref{eq:channel_coeff}) can be estimated, exception made for the additional losses and the phase misalignment, which are stochastic terms:
\begin{equation}
\label{eq:channel_lbmmse}
	\widetilde{h}_{i,n,s}^{(t)} = \frac{\widetilde{g}_{i,n,s}^{(TX,t)}\widetilde{g}_{i,n,s}^{(RX,t)}}{4\pi\frac{\widetilde{d}_{i,s}^{(t)}}{\lambda}\sqrt{\kappa B T_i}}e^{-\jmath\frac{2\pi}{\lambda}\widetilde{d}_{i,s}^{(t)}}
\end{equation}
where the tilde denotes that the terms are actually deduced from the relative positions of the UE and the swarm nodes. Clearly, the accuracy of the estimated channel coefficient is directly impacted by the accuracy of the location estimate performed by the UE and by the accuracy of the ephemeris data. In addition, as extensively discussed in Section~\ref{sec:assessment_ni}, $\widetilde{g}_{i,n,s}^{(TX,t)}$ and $\widetilde{g}_{i,n,s}^{(RX,t)}$ might also be impacted by a non-ideal knowledge of the radiation pattern, \emph{i.e.}, the antenna mathematical model used to compute the antenna pattern might not be flawlessly representing the actual radiation. From (\ref{eq:channel_lbmmse}), we can compute the beamforming matrix for the proposed \emph{Location-Based MMSE} (LB-MMSE) as:
\begin{equation}
\label{eq:lbmmse}
	 \mathbf{W}_{LB-MMSE}^{(t_0)} = {\widetilde{\mathbf{H}}}^H\left(\widetilde{\mathbf{H}}{\widetilde{\mathbf{H}}}^H + \mathrm{diag}(\boldsymbol{\alpha})I_{K}\right)^{-1}
\end{equation}
Compared to CSI-based techniques, LB-MMSE has the advantage of not requiring the UEs to estimate the channel coefficients, with manifold benefits. Firstly, the signalling overhead is significantly reduced, as the CSI vectors are not needed at the gNB-CU (OGC) or gNB-DU (OBC) and the UEs' positions require much smaller data packets to be transmitted\footnote{Assuming $32$ bits for floating point values, $64$ bits are needed per channel coefficient per user; this leads to $64\cdot N_F$ bits per user. For the location, each coordinate in the Global Positioning System (GPS) requires $24$ bits, leading to $48$ bits per user.}. Moreover, it shall be noticed that the estimation of the channel coefficients is typically assuming the presence of beams, each of which is associated to a known Data/Pilot sequence on which the estimation is performed; when moving to CF solutions, the concept of beam is lost and, thus, adjustments to the air interfaces would be needed. Finally, this approach also avoids the well-known issue of estimating the channel coefficients in low Signal-to-Interference Ratio (SIR) conditions. In fact, some of the known Data/Pilot sequences might be received with a significantly lower power compared to others. In such conditions, the UE is not able to estimate the channel coefficient, which shall then be either inferred at the gNB-CU/gNB-DU through more complex approaches, \emph{e.g.}, Machine Learning, or reported as a null in the channel matrix, which might lead to sparse ill-conditioned matrices.

\subsection{Beam-based MIMO}
\label{sec:model_beam}
Aiming at providing a complete performance comparison for the newly designed LB-MMSE algorithm, we also consider two beam-based solutions. The first one is a system implementing a \emph{phase-only} digital beam steering. In this case, an on-ground hexagonal beam lattice is defined based on the desired number of tiers around the center of the coverage area and on the 3 dB angular beamwidth, $\vartheta_{3dB}$, \cite{38821}. Denoting as $\mathbf{c}_{\ell,s}$ the $(u,v)$ coordinates of the generic $\ell$-th beam center from the $s$-th node\footnote{The $(u,v)$ system is centered at the satellite and, thus, the coordinates depend on the considered node, \cite{9142191}.}, its $n$-th beamforming coefficient is:
\begin{equation}
	b_{n,\ell,s} = \frac{1}{\sqrt{N_F}}e^{-\jmath k_0\mathbf{r}_{n,s} \cdot \mathbf{c}_{\ell,s}}
\end{equation}
where $k_0=2\pi/\lambda$ is the wave number and $\mathbf{r}_{n,s}$ the position of the $n$-th radiating element on the $s$-th on-board UPA. Assuming that in each time slot one user per beam is served, the beamforming matrix of the $s$-th node is fixed and it is given by $\mathbf{W}_{PO,s}={\left[b_{n,\ell,s}\right]}_{\substack{n=1,\ldots,N_F\\ \ell=1,\ldots,N_B}}$, where $N_B$ is the number of beams. The overall $\left(N_{node}N_F\right)\times N_B$ beamforming matrix $\mathbf{W}_{PO}$ is obtained by vertically concatenating the $N_F\times N_B$ $\mathbf{W}_{PO,s}$, $s=1,\ldots,N_{node}$, node matrices. It shall be noticed that, when advanced scheduling algorithms are implemented and not all of the beams are illuminated in all time slots, the Switchable Multi-Beam (MB) MIMO algorithm proposed in \cite{9142191} is obtained. In the following, we refer to MB beamforming also with all beams being illuminated.

The second beam-based approach is the SS-MMSE algorithm, \cite{eucnc_2022}. In this case, the generic $i$-th user is associated to the closest beam center, thus leading to a beam-based solution; then, the channel coefficients are estimated based on (\ref{eq:channel_lbmmse}), in which each term is computed based on the corresponding beam center location, and not the estimated UE location:
\begin{equation}
\label{eq:channel_ssmmse}
	\widetilde{h}_{i,n,s}^{(BC,t)} = \frac{\widetilde{g}_{i,n,s}^{(BC,TX,t)}\widetilde{g}_{i,n,s}^{(BC,RX,t)}}{4\pi\frac{\widetilde{d}_{i,s}^{(BC,t)}}{\lambda}\sqrt{\kappa B T_i}}e^{-\jmath\frac{2\pi}{\lambda}\widetilde{d}_{BC,i,s}^{(t)}}
\end{equation}
where $BC$ indicates that the terms shall be computed at the corresponding beam center. From this estimated channel matrix, the MMSE equation is again applied to obtain $\mathbf{W}_{SS-MMSE}$. Thus, this approach is different from LB-MMSE since is is beam-based and it approximates the UEs' locations to the those of closest beam centers.

Table~\ref{tab:algorithms} summarises the different beamforming techniques, reporting the required information, the beamforming and channel coefficient equations, and the potential sources of misalignment between the channel matrices at $t_0$ and $t_1$. It shall be noticed that, for SS-MMSE, $\widetilde{\mathbf{H}}_{BC}$ denotes the beam centers channel matrix estimated based on (\ref{eq:channel_ssmmse}).

\subsection{Power normalisation}
\label{sec:model_norm}
The normalisation of the beamforming matrix is a fundamental operation, as extensively discussed in \cite{eucnc_2022,9438169}. In fact, the Frobenius norm of the beamforming matrix, ${\left\|\mathbf{W}\right\|}_F^2$, represents the total emitted power and, by applying the MMSE formula, there is no guarantee that such power will be upper-bounded so as to not exceed the total available power. First considering a single node in the swarm, for the sake of simplicity, this means that there might be situations in which ${\left\|\mathbf{W}\right\|}_F^2>P_t$, where $P_t$ denotes the total available power on-board. Aiming at addressing this issue, several normalisations have been considered in the literature:
\begin{itemize}
	\item Sum Power Constraint (SPC):
	\begin{equation}
		\widetilde{\mathbf{W}} = \sqrt{\frac{P_t}{{\left\|\mathbf{W}\right\|}_F^2}}\mathbf{W} = \sqrt{\frac{P_t}{\mathrm{tr}\left(\mathbf{W}\mathbf{W}^H\right)}}\mathbf{W}
	\end{equation}
	This normalisation guarantees that: i) the overall power allocated by the beamforming matrix is equal to that actually available, \emph{i.e.}, ${\left\|\mathbf{W}\right\|}_F^2=P_t$; ii) preserves the orthogonality among the beamforming matrix columns, \emph{i.e.}, it does not disrupt the optimal MMSE solution since all columns are normalised by the same scalar quantity. However, SPC does not control the power emitted per radiating element, which might lead to a performance degradation due to driving the on-board High Power Amplifiers (HPAs) close to or above the saturation level, thus introducing undesired non-linear effects.
	\item Maximum Power Constraint (MPC)
	\begin{equation}
	\label{eq:MPC}
		\widetilde{\mathbf{W}} = \sqrt{\frac{P_t}{K\max_{k=1,\ldots,K}{\left\|\mathbf{w}_{:,k} \right\|}^2}}\mathbf{W} 
	\end{equation}
	In this case, the normalisation is similar to SPC, with the only difference being that only one radiating element is emitting the maximum allowed power, while all of the others emit a lower power level. Thus: i) the overall emitted power is below the available power, \emph{i.e.}, ${\left\|\mathbf{W}\right\|}_F^2<P_t$; ii) the orthogonality among the beamforming matrix columns is preserved; and iii) the emitted power is limited per radiating element. However, since only one radiating element is emitting the maximum power, while the others are significantly limited in their emissions, this approach might lead to a degradation of the SNR and, in general, to a performance loss.
	\item Per Antenna Constraint (PAC)
	\begin{equation}
	\label{eq:PAC}
		\widetilde{\mathbf{W}} = \sqrt{\frac{P_t}{K}}\mathrm{diag}\left(\frac{1}{\left\|\mathbf{w}_{1,:}\right\|},\ldots, \frac{1}{\left\|\mathbf{w}_{K,:}\right\|}\right)\mathbf{W} 
	\end{equation}
	With this approach, each radiating element transmits at an equal power level, thus ensuring that the overall available power is not exceeded, \emph{i.e.}, ${\left\|\mathbf{W}\right\|}_F^2=P_t$. However, since each row of the beamforming matrix is normalised independently from each other, the orthogonality of the beamformer columns are disrupted and potentially large performance degradation is introduced.
\end{itemize}
When multiple NGSO nodes are considered, some further considerations are needed. In this case, the beamforming matrix includes the power emitted by radiating elements on-board different nodes. Consequently, the normalisations shall be adjusted to satisfy the power constraints per node. Assuming that each node has the same available on-board power, $P_{t,node}$, we introduce the following normalisations:
\begin{itemize}
	\item swarm SPC (sSPC): directly applying the SPC normalisation with a total power $N_{node}P_{t,node}$ guarantees that this power is not exceeded at swarm level, \emph{i.e.}, ${\left\|\mathbf{W}\right\|}_F^2=N_{node}P_{t,node}$; however, a single node might be required to emit more power than available. To circumvent this issue, the swarm-based sSPC normalisation is introduced based on the observation that the overall $\left(N_{node}N_{F}\right)\times K$ beamforming matrix can be divided in blocks corresponding to the single nodes beamforming matrices, \emph{i.e.}:
	\begin{equation}
		\mathbf{W} = \begin{bmatrix}
		\mathbf{W}_1\\
		\vdots \\
		\mathbf{W}_{N_{node}}
		\end{bmatrix}
	\end{equation}
		with $\mathbf{W}_s$ denoting the $N_F\times K$ beamforming matrix of the $s$-th NGSO node. Each node beamforming matrix can be normalised with the SPC approach as a standalone matrix, guaranteeing that: i) the overall emitted power satisfies ${\left\|\mathbf{W}\right\|}_F^2=N_{node}P_{t,node}$; and ii) each satellite emits a power ${\left\|\mathbf{W}_s\right\|}_F^2=P_{t,node}$, $s=1,\ldots,N_{node}$. Clearly, this approach leads to a slight degradation in the performance, because a normalisation that is not scalar for the entire beamforming matrix $\mathbf{W}$ leads to a loss of orthogonality in the beamforming matrix columns, \emph{i.e.}, to a loss in its  interference cancellation capabilities. 
	\item swarm MPC (sMPC): in this case, there are two options. In fact, if the objective is to preserve the orthogonality in the beamforming matrix columns, then (\ref{eq:MPC}) can be directly applied since only one radiating element will emit its maximum power; this guarantees the preservation of the orthogonality, but actually leads to lower emitted power levels, since only a single element from a single node in the swarm will transmit the maximum power. Another possibility is to better exploit the available power, by normalising with the MPC approach each node matrix, as in the sSPC solution; this guarantees that the overall emitted power is still satisfying the condition ${\left\|\mathbf{W}\right\|}_F^2<N_{node}P_{t,node}$ and that each node emits a power ${\left\|\mathbf{W}_s\right\|}_F^2<P_{t,node}$, $s=1,\ldots,N_{node}$. Only a single element per node emits its maximum power. In the following, we assume the latter solution is implemented.
\end{itemize}
As for PAC, from (\ref{eq:PAC}) it can be noticed that no issue arises since each row of the beamforming matrix is individually normalised to guarantee that each radiating element emits the same power level. Thus, as previously discussed, we are ensuring that each nodes emits a total power $P_{t,node}$ and that the swarm collectively emits a power $N_{node}P_{t,node}$, with a performance loss related to the disrupted orthogonality in the beamforming matrix columns on a node basis.

In the following, we consider the (s)SPC and (s)MPC normalisations. It was observed that, as expected, the PAC approach has a poor interference cancellation performance and, as such, it usually leads to very low SINRs. The (s)MPC solution is a viable normalisation approach since it satisfies all power constraints and avoids operating the on-board HPAs in the non-linear region. The (s)SPC normalisation is retained as it provides an upper-bound performance, even though not practically implementable due to the possibility that a node is required to transmit more power than available.

\begin{table}[t!]
\renewcommand{\arraystretch}{1.3}
\centering
\caption{System configuration parameters for the numerical assessment.}
\label{tab:parameters}
 \begin{tabular}{|c|c|} 
 \hline
 \textbf{Parameter} & \textbf{Value} \\
 \hline
 Transmission EIRP & $0, 4, 8, 12$ dBW/MHz \\
 \hline
 Channel model & clear-sky, NLOS \\
 \hline
 Propagation environment & dense-urban \\
 \hline
Carrier frequency & S-band ($2$ GHz) \\
\hline
User bandwidth & $30$ MHz\\
\hline
Receiver type & \makecell[c]{fixed VSAT/handheld \\ (antenna parameters in \cite{38821})} \\
\hline
$N_F$ & \makecell[c]{$1024$ ($32\times 32$ UPA)\\ $0.55\lambda$ spacing} \\
\hline
Nodes per swarm $N_{node}$ & $1,2$ \\
\hline
Altitude & $600$ km \\
\hline
Beams per node $N_{B,node}$ & \makecell[c]{$91 (N_{node}=1)$ \\ $61 (N_{node}=2)$} \\
\hline
User density & $0.5$ users/km$^2$ \\
\hline
Aging interval $\Delta t$ & $16.7$ ms \\
\hline
Scheduling & random \\
\hline
 \end{tabular}
\end{table}

\section{Numerical Assessment}
\label{sec:assessment}
In this Section, we discuss the extensive numerical results obtained through Monte Carlo simulations with the MMSE, LB-MMSE, SS-MMSE, and MB algorithms. The system configuration parameters for the numerical assessment are reported in Table~\ref{tab:parameters}. It is worthwhile noticing that we consider with $\Delta t = 16.7$ ms, computed based on OGC; clearly, lower values will results in a better performance, while larger values in worse results due to the reduced/increased information aging at the transmitter. However, the general trends discussed below still hold. Moreover, for the following system-level analyses, only the value of $\Delta t$ impacts the results and not the specific choice of a OGC or OBC design.

\subsection{Simulation scenarios, assumptions, and metrics}
The users are uniformly distributed in the coverage area. Based on the user density reported in Table~\ref{tab:parameters}, approximately $30000$ users are considered. The UEs are assumed to be fixed; in fact, by means of extensive numerical simulations, it was observed that for users moving at up to $250$ km/h the performance loss was in the order of $10^{-4}$  bit/s/Hz in terms of spectral efficiency, \emph{i.e.}, negligible. To provide connectivity to the uniformly distributed users, the UPA on-board each node is designed as a square lattice with $N_F=1024$ radiating elements; the spacing between adjacent elements is fixed at $0.55\lambda$. Notably, the radiation pattern from the $n$-th element on the $s$-th node in the direction of the $i$-th user, identified by the direction cosines coordinates $\mathbf{p}_{i,s}^{(t)}=\left[u_{i,s}^{(t)},v_{i,s}^{(t)}\right]$ at the $t$-th time instant, is obtained as, \cite{9142191}:
	\begin{equation}
		g^{(TX,t)}_{i,n,s} = g_{E,i}^{(t)}e^{\jmath k_0\mathbf{r}_{n,s} \cdot \mathbf{p}_{i,s}^{(t)}}
	\end{equation}
	where $g_{E,i}$ is the radiation pattern of the single radiating element in the direction of the $i$-th user $\mathbf{r}_{n,s}$ the position of the $n$-th element on the $s$-th UPA. The element pattern $g_{E,i}$ is assumed to be the same for all radiating elements and computed as in \cite[Section 5.1]{itu_antenna}.
	
\begin{figure}[t!]
	\centering
	\includegraphics[width=\columnwidth]{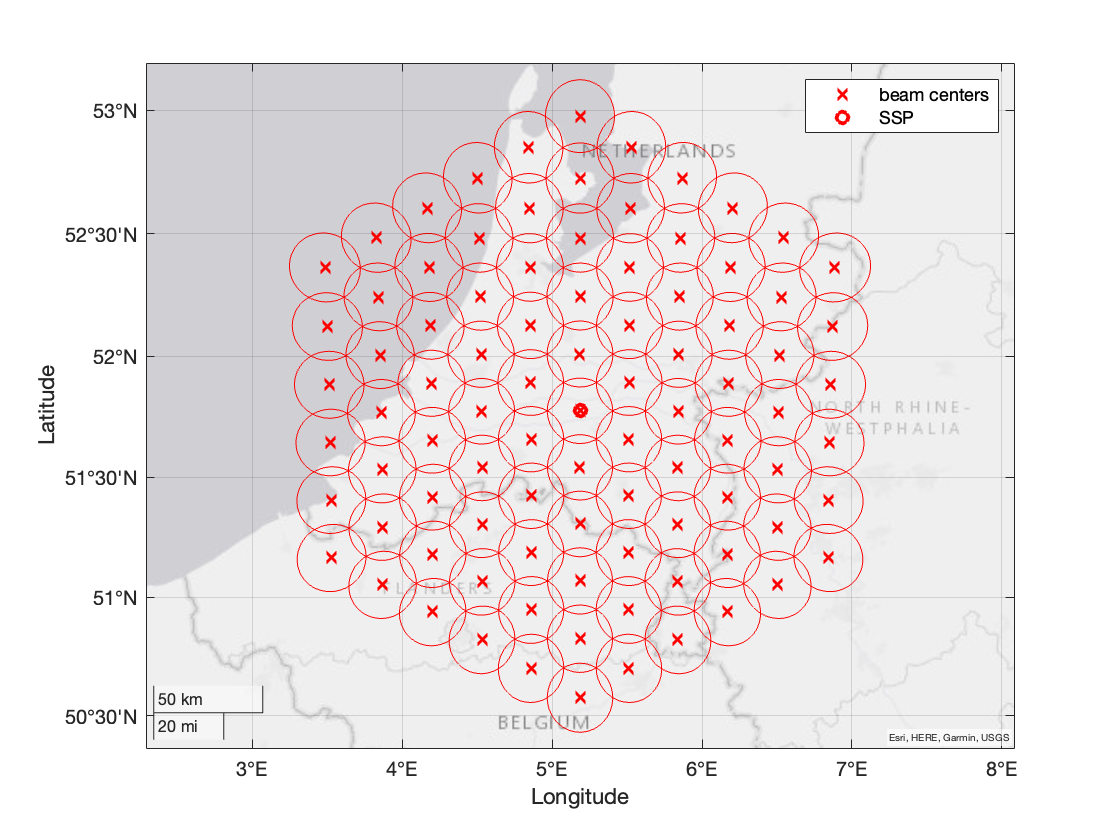}
	\caption{Centralised ($N_{node}=1$) scenario.}
	\label{fig:scenario_n1}
\end{figure}

\begin{figure}[t!]
	\centering
	\includegraphics[width=\columnwidth]{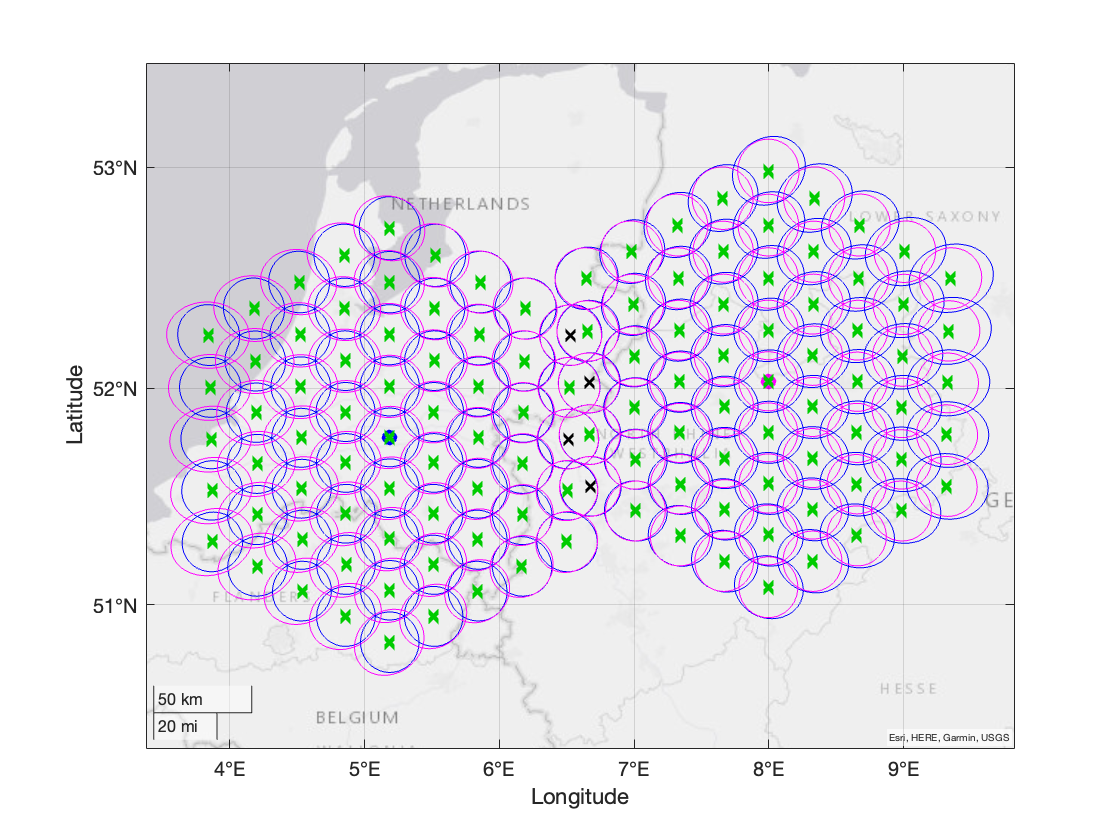}
	\caption{Example of RRM algorithm activating a subset of beams (green) to avoid an ill-conditioned channel matrix due to beam proximity issues.}
	\label{fig:active_beams}
\end{figure}

We consider both a centralised ($N_{node}=1$) and federated ($N_{node}=2$)  configurations. In order to provide a fair comparison between Cell-Free (MMSE, LB-MMSE) and beam-based (SS-MMSE, MB) algorithms, each node in the swarm generates $N_{B,node}$ beams organised in an hexagonal lattice on-ground. The number of beams per node is different in the two scenarios so as to obtain a similar number of beams per swarm $N_B$ ($91$ with $N_{node}=1$ and $122$ with $N_{node}=2$). Moreover, it shall be noticed that, with multiple nodes, each node in the swarm generates its corresponding lattice, but then it covers all of the beams created by the swarm. Figures \ref{fig:scenario_n1} and \ref{fig:scenario_n2} show the on-ground beam footprints with $N_{node}=1$ and $N_{node}=2$, respectively. With multiple nodes, the fact that each node is initially generating a single beam lattice leads to beams that significantly overlap at the border between the two lattices, \emph{i.e.}, there are beams that have their centers inside other beams boundaries at less than $-3$ dB. Notably, when implementing MIMO techniques, this might lead to a performance loss, \cite{9438169,8510728}; in particular, two UEs selected from two of such beams might be scheduled in the same time slot. In this case, the two users have very similar channel signatures (CSI coefficients) and, thus, the channel matrix to be inverted in MMSE-like solutions might be ill-conditioned. To circumvent this issue, proper scheduling algorithms might be implemented; in the following, we assume that the RRM algorithm avoids such situations by activating only one beam among those in which the relative distance among the beam centers does not guarantee a $3$ dB separation. An example is shown in Figure~\ref{fig:active_beams}. Consequently, to provide a fair comparison between the centralised and federated scenarios, the available power per node is scaled to guarantee that the same average power per beam is available:
\begin{figure}[t!]
	\centering
	\includegraphics[width=\columnwidth]{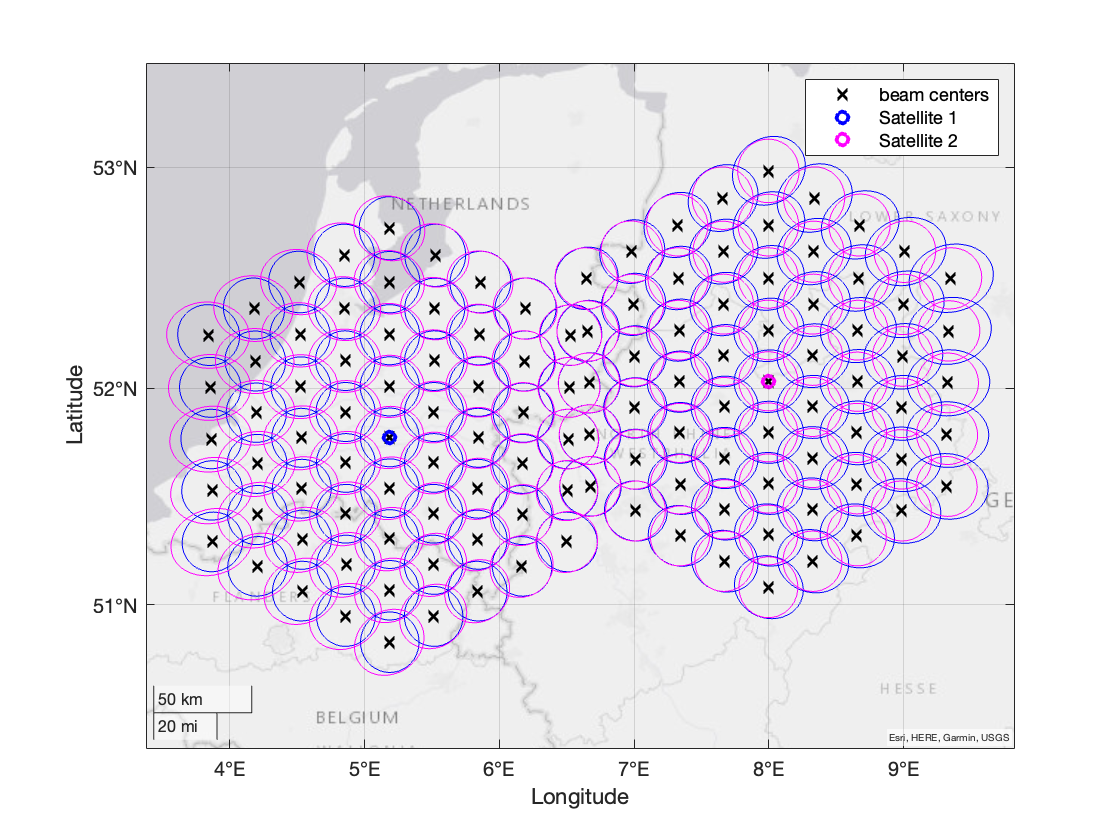}
	\caption{Federated ($N_{node}=2$) scenario. Blue lines represent the footprint of satellite 1 and magenta lines those for satellite 2.}
	\label{fig:scenario_n2}
\end{figure}
	\begin{equation}
		P_{t,node} = P_{t}\frac{N_{B,multi}}{N_{node}N_{B,single}}
	\end{equation}
	where $P_{t,node}$ is the available power per node and $N_{B,single}$, $N_{B,multi}$ the number of active beams with $N_{node}=1$ and $N_{node}=2$, respectively. It can be noticed that when all beams are active in the multiple nodes case (\emph{i.e.}, there is no beam proximity issue): i) if $N_{B,multi}=N_{node}N_{B,single}$, \emph{i.e.} each node in the swarm generates the same amount of beams as in the centralised case, then $P_{t,node}=P_{t}$; and ii) if $N_{B,multi}=N_{B,single}$, \emph{i.e.} the same number of beams is generated independently of the number of nodes, $P_{t,node}=P_{t}/N_{node}$. Referring to Figure~\ref{fig:scenario_n2} and \ref{fig:active_beams}, there are $118$ active beams compared to $91$ with a single node; a transmission power density of $4$ dBW/MHz leads to $P_t = 18.77$ dBW over a $30$ MHz bandwidth and to $P_{t,node} = 16.89$ dBW/MHz.

\begin{figure*}
    \centering
    \subfigure[$N_{node}=1$]
    {
        \includegraphics[width=0.48\textwidth]{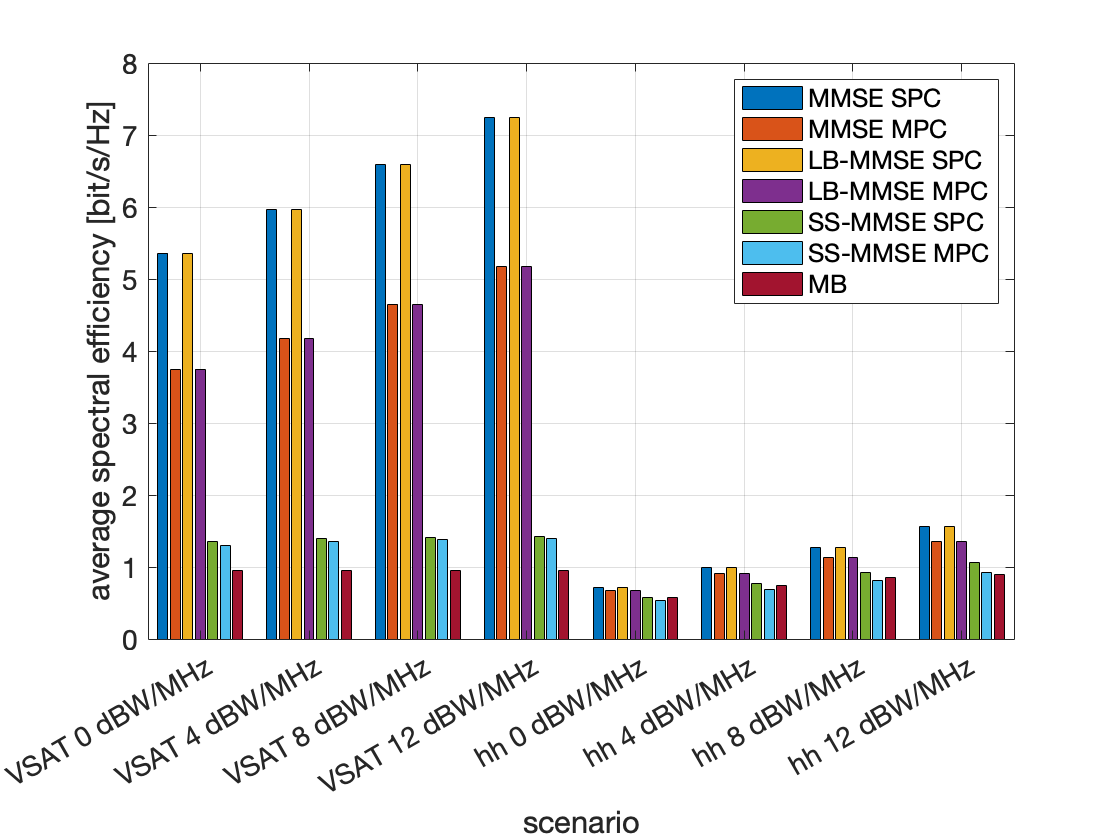}
        \label{fig:single_tlos_rate}
    }
    \subfigure[$N_{node}=2$]
    {
        \includegraphics[width=0.48\textwidth]{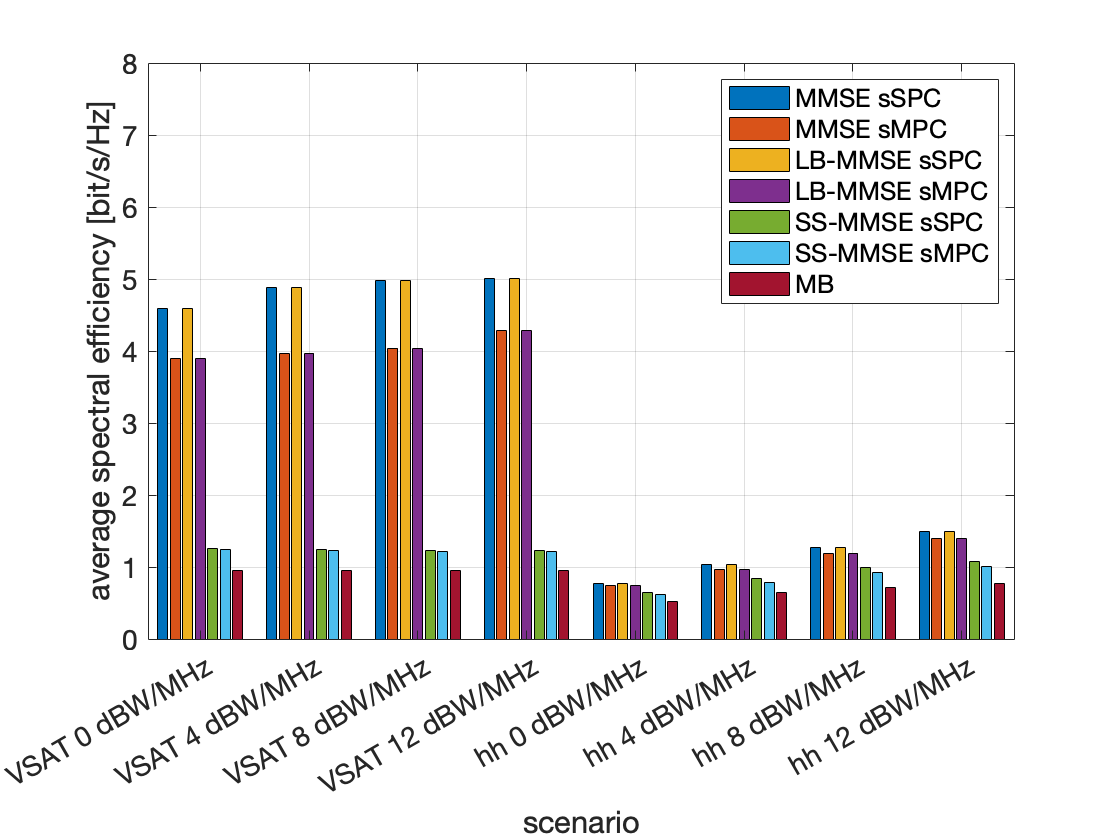}
        \label{fig:dual_tlos_rate}
    }
    \caption{Average spectral efficiency in clear-sky with $N_{node}=1$ (left) and $N_{node}=2$ (right).}
    \label{fig:tlos_rate}
\end{figure*}
\begin{figure*}
    \centering
    \subfigure[$N_{node}=1$]
    {
        \includegraphics[width=0.48\textwidth]{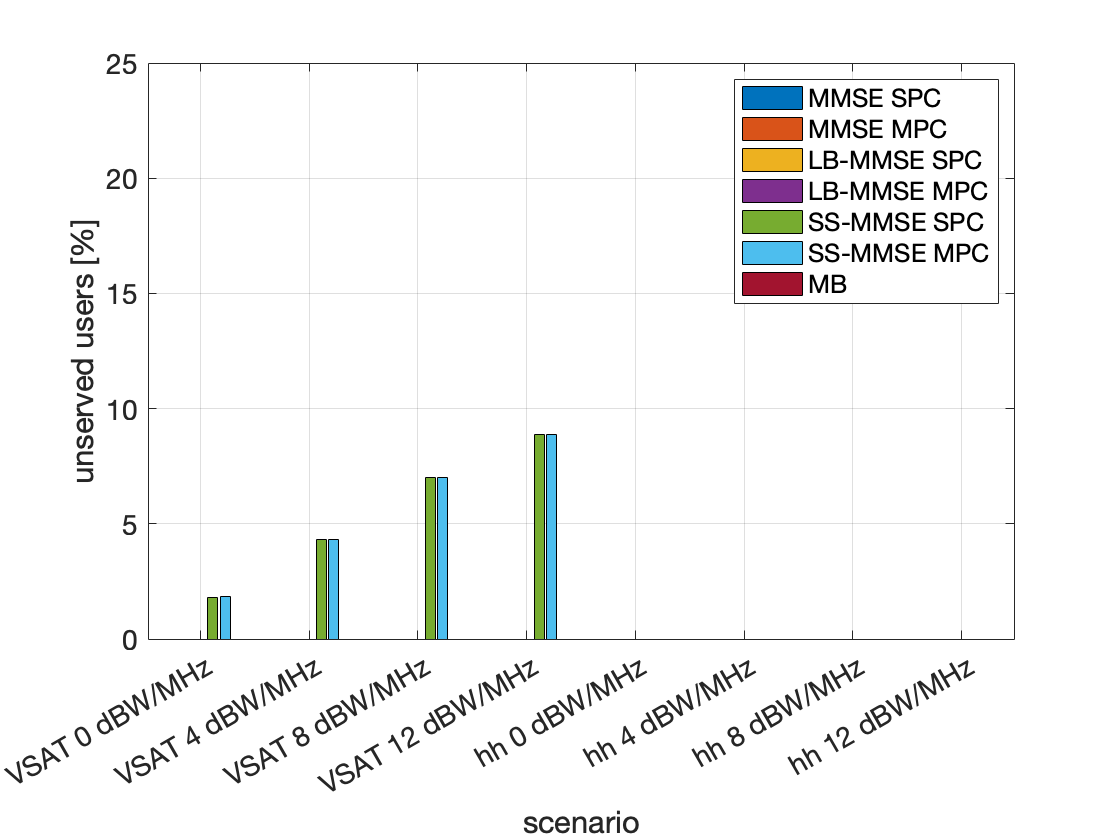}
        \label{fig:single_tlos_rate}
    }
    \subfigure[$N_{node}=2$]
    {
        \includegraphics[width=0.48\textwidth]{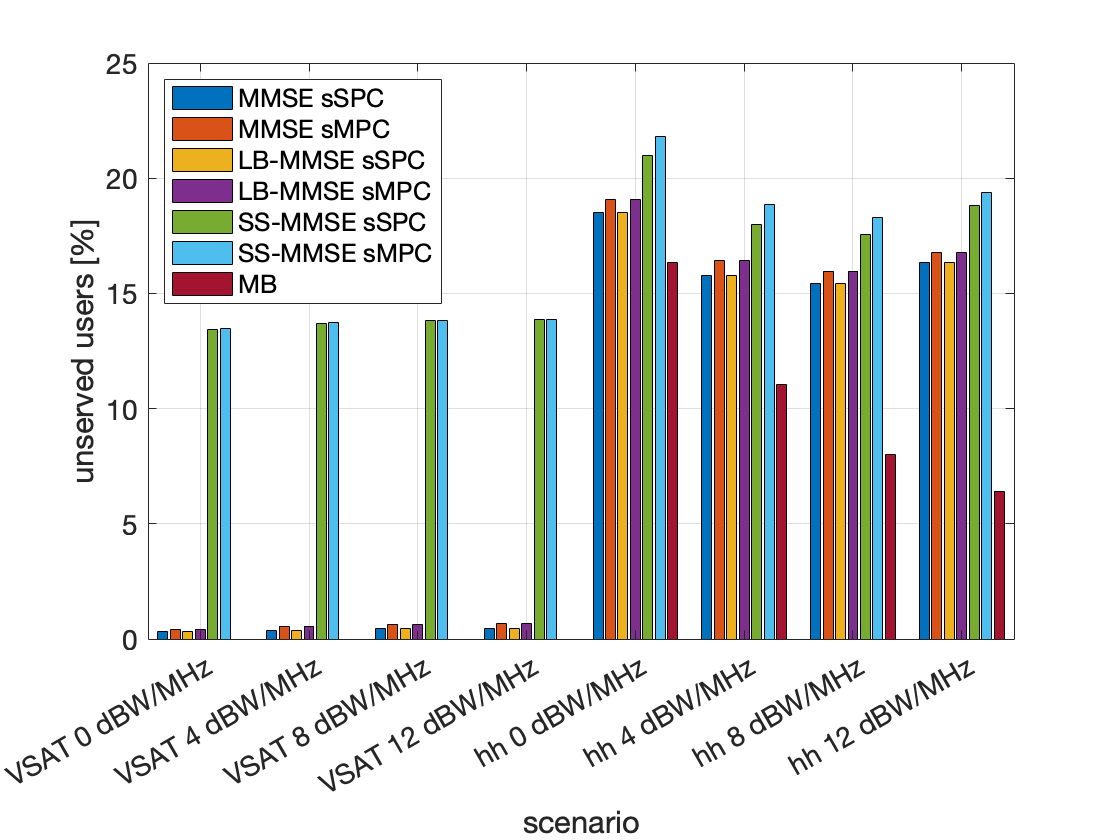}
        \label{fig:dual_tlos_rate}
    }
    \caption{Percentage of unserved users in clear-sky with $N_{node}=1$ (left) and $N_{node}=2$ (right).}
    \label{fig:tlos_out}
\end{figure*}

With respect to scheduling, a random algorithm is implemented, \cite{eucnc_2022,9438169}: in each time slot, one UE per beam is randomly selected and the total number of time slots is computed so as to guarantee that all UEs are served at least once. The coverage is based on Earth moving beams, \emph{i.e.}, the beams move together with the NGSO nodes along their orbits. It is worthwhile highlighting that a beam lattice is only needed for the MB and SS-MMSE solutions; MMSE and LB-MMSE allow to implement CF-MIMO, as long as an advanced scheduler operating exlusively on the location or CSI vectors is implemented. Thus, the concept of beams for MMSE and LB-MMSE is only exploited for the random scheduler and to have a comparison with MB and SS-MMSE.

Two channels are considered: i) clear-sky, in which no additional loss is present, \emph{i.e.}, $L_{i,s}^{(t)}=1$, $\forall i,s,t$; and ii) NLOS, modelled as described in Section~\ref{sec:model_beam} assuming a dense-urban environment, \emph{i.e.}, the worst conditions in terms of clutter loss and $\sigma_{SHA}$. In the latter scenario each UE is in LOS or NLOS conditions according to the probabilities reported in \cite{38811} as a function of its elevation angle.

Below, we discuss the numerical results obtained with both ideal and non-ideal information to compute the beamforming matrix. The performance is provided in terms of average spectral efficiency in [bit/s/Hz] and percentage of unserved users. To this aim, it shall be mentioned that, based on the truncated Shannon bound in  (\ref{eq:shannon_trunc}), the average spectral efficiency is computed only on the UEs which are served by the system, \emph{i.e.}, those for which the SINR is above $\gamma_{min}$. The percentage of unserved users is computed as the percentage of users with SINR below $\gamma_{min}$.

\subsection{Ideal information}
In this Section, we discuss the numerical results under ideal conditions, \emph{i.e.}, ideal CSI estimation (MMSE), ideal location estimation (LB-MMSE, SS-MMSE, MB), ideal knowledge of the radiation pattern (LB-MMSE and SS-MMSE). It is worthwhile mentioning that, despite the ideal conditions, there is still a misalignment between the beamforming matrix $\mathbf{W}^{(t_0)}$ and the channel matrix $\mathbf{H}^{(t_1)}$ because of the movement of the swarm nodes on their orbits.

Figure~\ref{fig:tlos_rate} shows the average spectral efficiency in clear-sky conditions with $N_{node}=1$ and $N_{node}=2$. The following general trends can be observed:
\begin{itemize}
	\item MMSE and LB-MMSE provide the best performance for all transmission power densities and terminal types. Moreover, they provide the same performance; this is motivated by observing that, in the absence of additional losses and with ideal estimations, the channel coefficients in (\ref{eq:channel_coeff}) and (\ref{eq:channel_lbmmse}) are identical. The performance of the SS-MMSE and MB solutions are significantly worse, with losses in the order of $6$ and $4$ bit/s/Hz with VSATs in the centralised and federated scenarios.
	\item As expected, the (s)SPC normalisation provides the best performance. However, this solution provides a theoretical upper-bound but it is not feasible. With (s)MPC, the performance is worse in particular in systems with large received power (VSAT and large power density); this is motivated by the fact that the (s)MPC normalisation does not exploit the entire available power.
	\item Handheld terminals have a worse performance compared to VSATs, due to the omni-directional antennas. The different techniques and normalisations provide a similar performance with low EIRP (in this case, the performance is more noise limited than interference limited), while with larger transmission power levels interference increases and the impact of a different technique or normalisation becomes slightly more evident.
\end{itemize}

\begin{figure*}
    \centering
    \subfigure[SNR]
    {
        \includegraphics[width=0.48\textwidth]{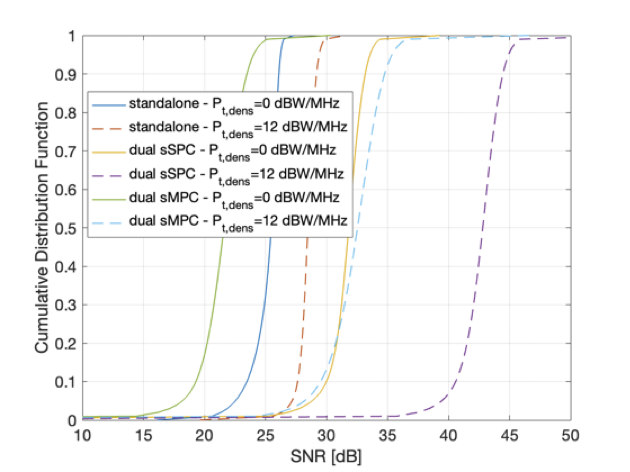}
        \label{fig:cdf_tlos_snr}
    }
    \subfigure[SIR]
    {
        \includegraphics[width=0.48\textwidth]{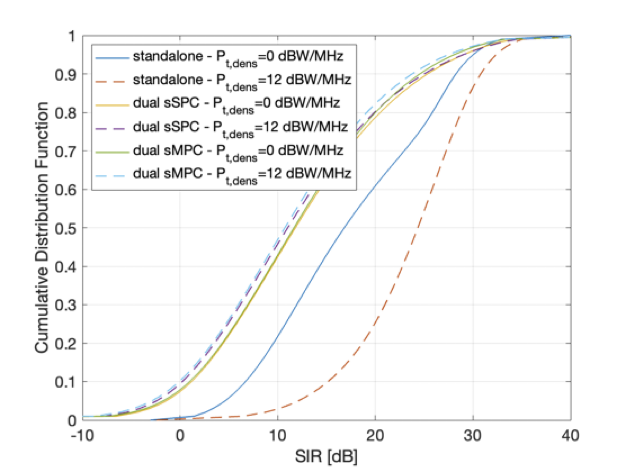}
        \label{fig:cdf_tlos_sir}
    }
    \caption{CDF of the SNR (left) and SIR (right) with MMSE beamforming and VSATs. For the centralised case, SPC is considered.}
    \label{fig:cdf_tlos}
\end{figure*}
\begin{figure*}
    \centering
    \subfigure[$N_{node}=1$]
    {
        \includegraphics[width=0.48\textwidth]{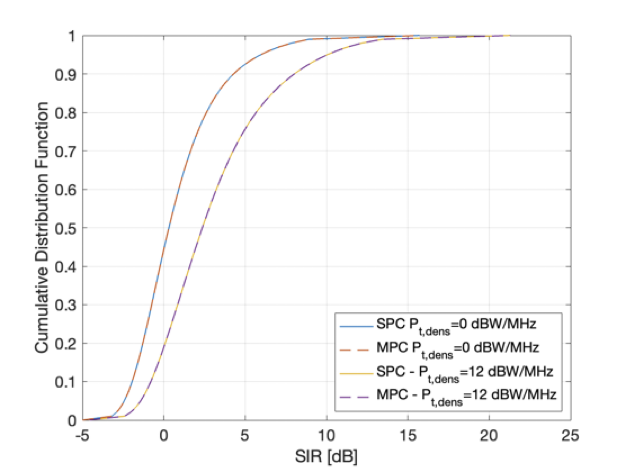}
        \label{fig:cdf_tlos_snr}
    }
    \subfigure[$N_{node}=2$]
    {
        \includegraphics[width=0.48\textwidth]{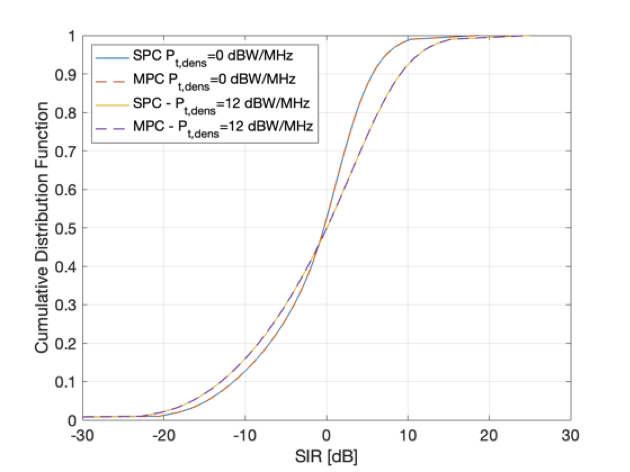}
        \label{fig:cdf_tlos_sir_hh}
    }
    \caption{CDF of the SIR in the centralised (left) and federated (right) scenarios with MMSE beamforming and handheld terminals.}
    \label{fig:cdf_tlos_hh}
\end{figure*}

Comparing the performance between the centralised and federated scenarios, interestingly, it can be noticed that:
\begin{figure*}
	\centering
	\subfigure[$N_{node}=1$]
	{
		\includegraphics[width=0.48\textwidth]{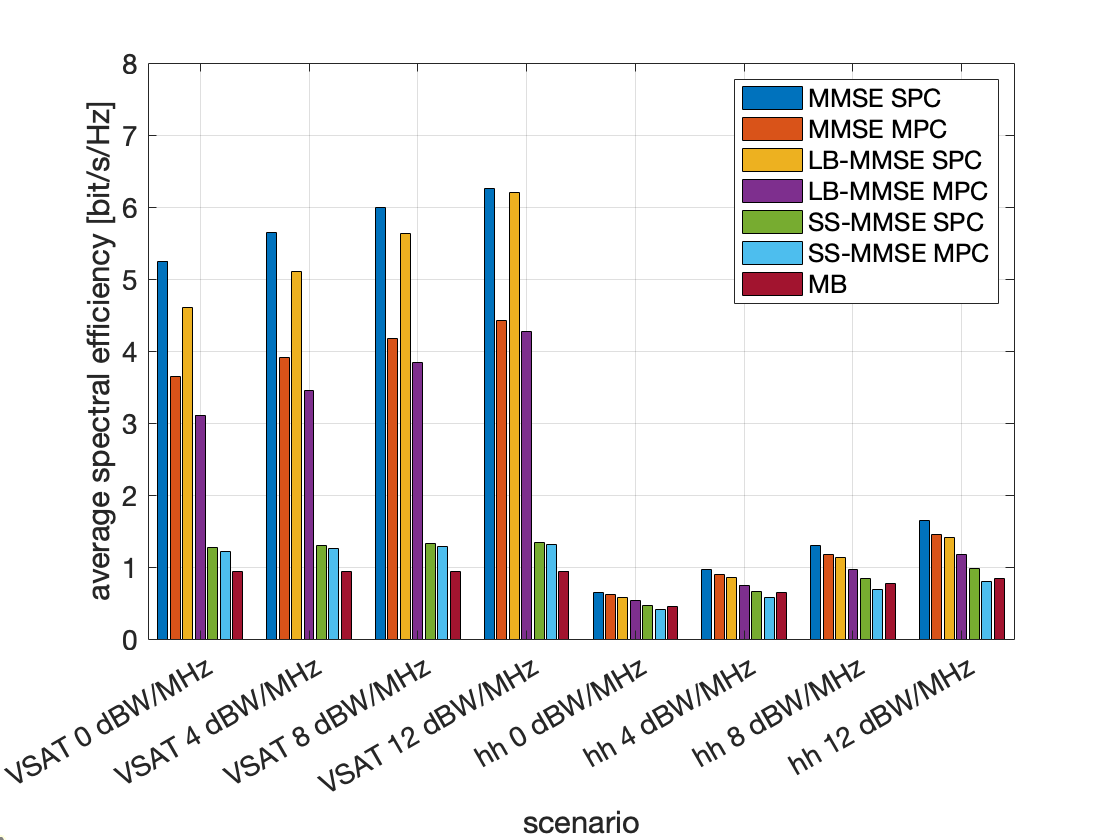}
		\label{fig:single_nlos_rate}
	}
	\subfigure[$N_{node}=2$]
	{
		\includegraphics[width=0.48\textwidth]{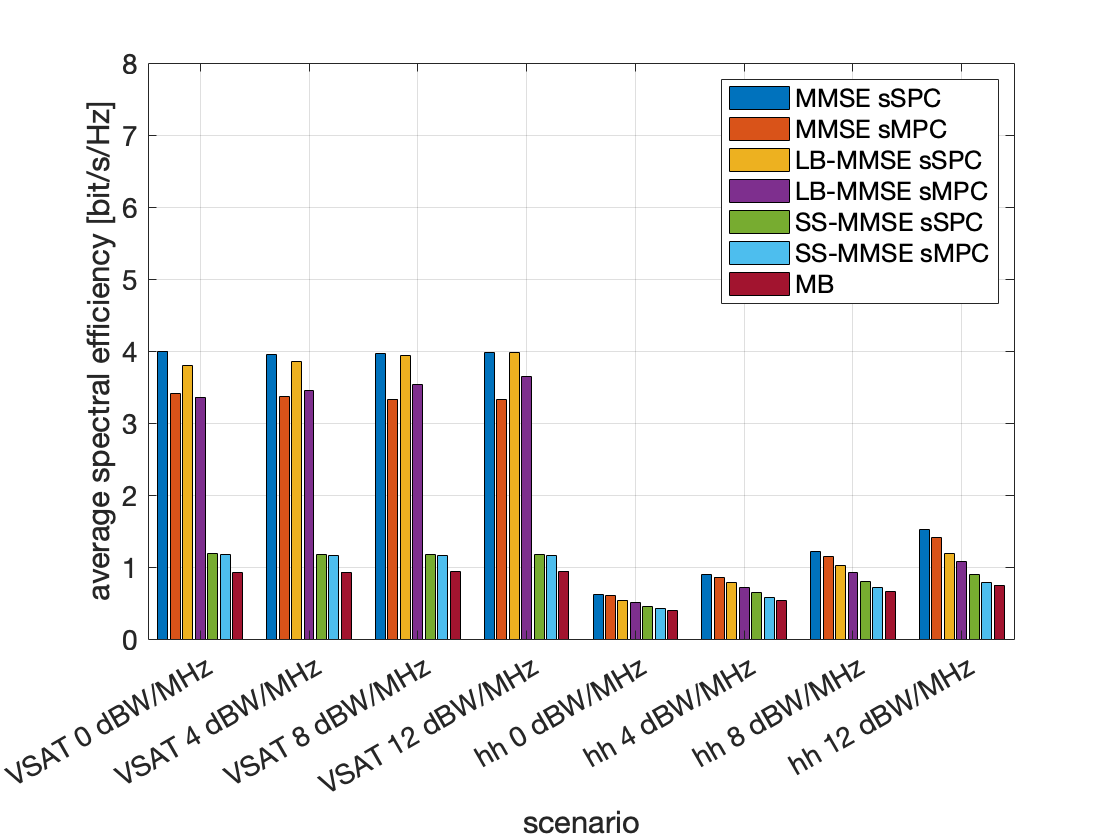}
		\label{fig:dual_nlos_rate}
	}
	\caption{Average spectral efficiency in NLOS dense-urban conditions with $N_{node}=1$ (left) and $N_{node}=2$ (right).}
	\label{fig:nlos_rate}
\end{figure*}
\begin{figure*}
	\centering
	\subfigure[$N_{node}=1$]
	{
		\includegraphics[width=0.48\textwidth]{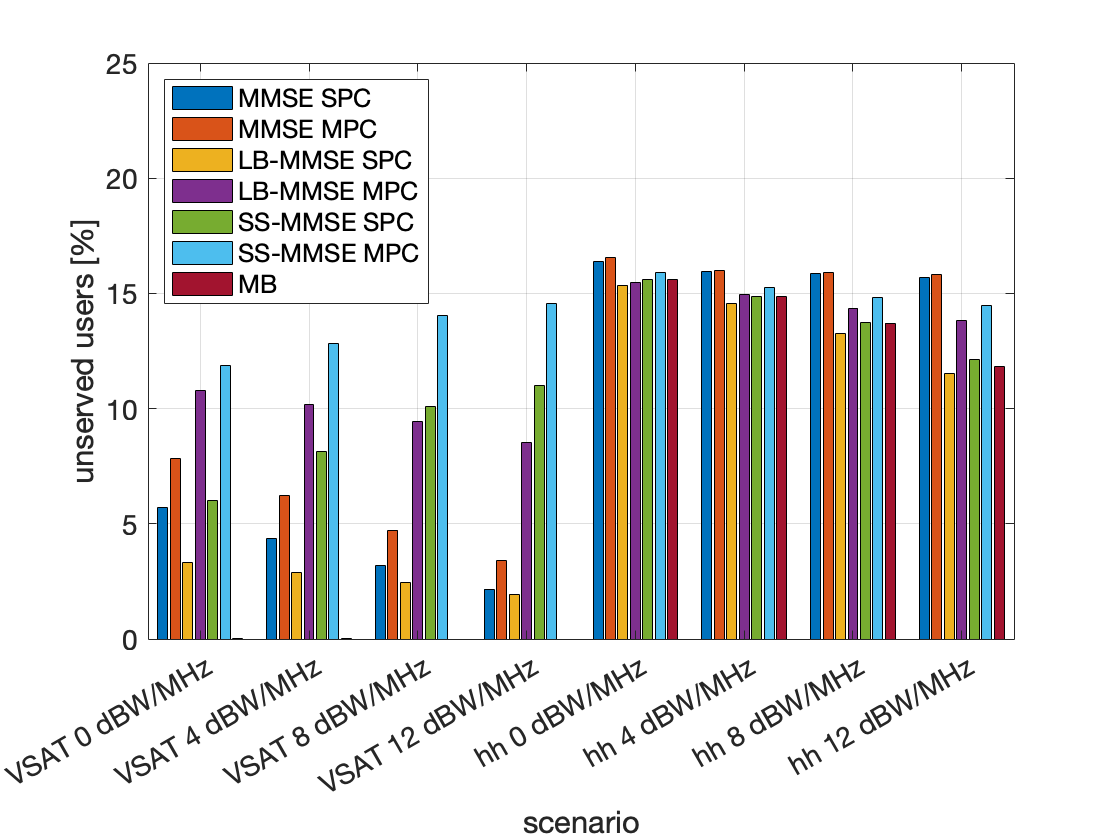}
		\label{fig:single_nlos_rate}
	}
	\subfigure[$N_{node}=2$]
	{
		\includegraphics[width=0.48\textwidth]{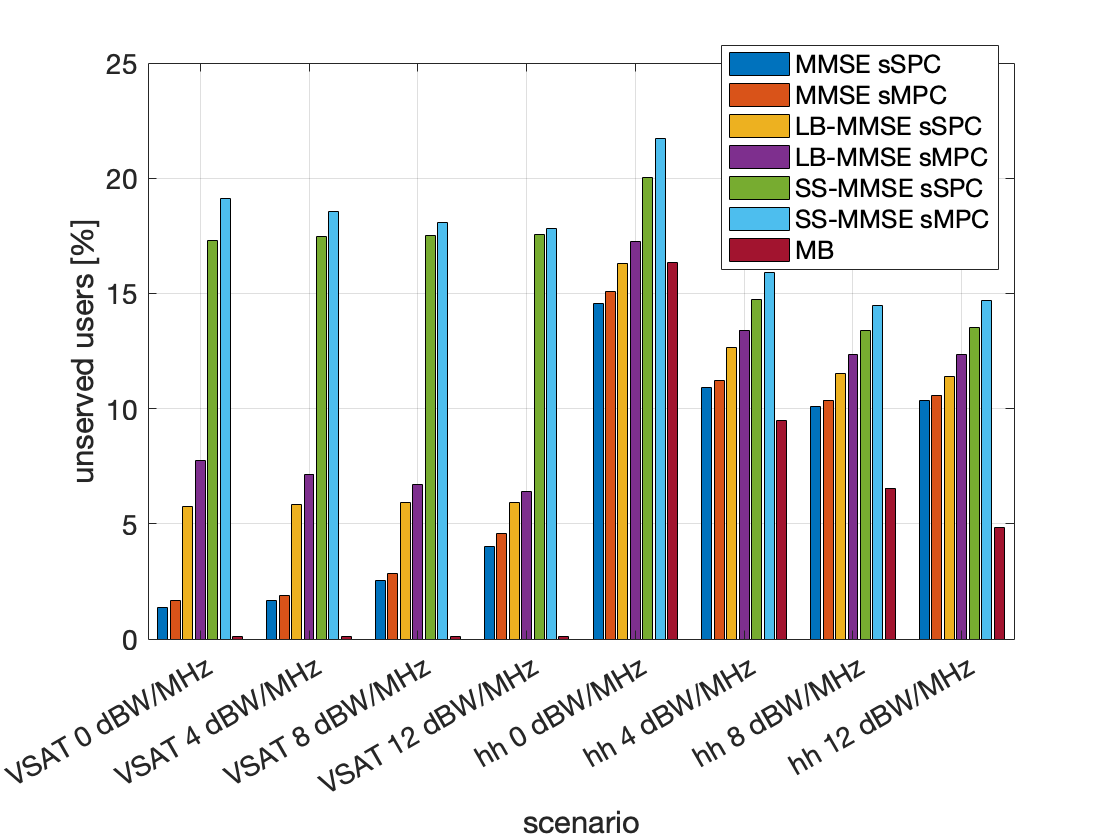}
		\label{fig:dual_nlos_rate}
	}
	\caption{Percentage of unserved users in NLOS dense-urban conditions with $N_{node}=1$ (left) and $N_{node}=2$ (right).}
	\label{fig:nlos_out}
\end{figure*}
\begin{itemize}
	\item With both handheld and VSAT terminals, the spectral efficiency is larger for increasing values of the power density. However, with multiple nodes and VSATs, the performance tends to saturate with large power levels. Figure~\ref{fig:cdf_tlos} shows the CDF of the SNR and SIR with MMSE beamforming. It can be observed that, clearly, the SNR is always improved by increasing the power density; however, the SIR is slightly worse with larger power with $N_{node}=2$, while with $N_{node}=1$ it is improved. Thus, with multiple nodes in the swarm interference cancellation is critical. This is motivated by observing that, with $N_{node}$ nodes there are $N_{nodes}N_F$ radiating elements; this leads to an increased sensitivity to any misalignment between $\mathbf{W}^{(t_0)}$ and $\mathbf{H}^{(t_1)}$. Moreover, also the block sSPC and sMPC normalisations with multiple nodes increase such misalignment. 
	\item The performance with VSATs and $N_{node}=2$ is beneficial only at low power and with (s)MPC. In all other cases, \emph{i.e.}, when the interference to be dealt with by the beamformer is increased, the centralised scenario provides a better result.
	\item With handheld terminals, the average spectral efficiency is typically slightly better ($0.05-0.1$ bit/s/Hz) with $N_{node}=2$. The only exception is when the power density is large and (s)SPC normalisations are considered. However, in this case we should also observe the percentage of unserved users, reported in Figure~\ref{fig:tlos_out}. In the centralised case, there are no users in outage with MMSE and LB-MMSE; with SS-MMSE, based on an approximation of the UEs' locations, the impact of a larger transmission power is more detrimental for the interference cancellation capability and the outage increases. With two nodes, the situation is worse. In particular:
\begin{itemize}
		\item with SS-MMSE there are many unserved users with both terminal types (up to $22\%$). This is motivated by the much larger number of radiating elements and the consequent impact on any misalignment between $\mathbf{W}^{(t_0)}$ and $\mathbf{H}^{(t_1)}$, more evident in this case due to the further approximation in the UEs' locations;
		\item there is a slight outage also with MMSE/LB-MMSE and VSATs, always below $0.7\%$, motivated as above. With VSATs, the impact of such misalignment is limited by the presence of directive antennas;
		\item with MMSE/LB-MMSE and handheld terminals, the outage is significantly higher (up to $19\%$ in some cases) and it tends to decrease for increasing power levels. This trend is motivated by the omni-directional antennas at the receiver. In fact, VSATs are assumed to be ideally pointing and tracking the node providing the best radiation pattern, thus limiting the residual intra-swarm interference. Handheld terminals do not have directive radiation patterns and, thus, they are significantly more subject to the residual intra-swarm interference. This behaviour is shown in Figure~\ref{fig:cdf_tlos_hh}.
	\end{itemize}
\end{itemize}
\begin{figure*}
    \centering
    \subfigure[$N_{node}=1$]
    {
        \includegraphics[width=0.48\textwidth]{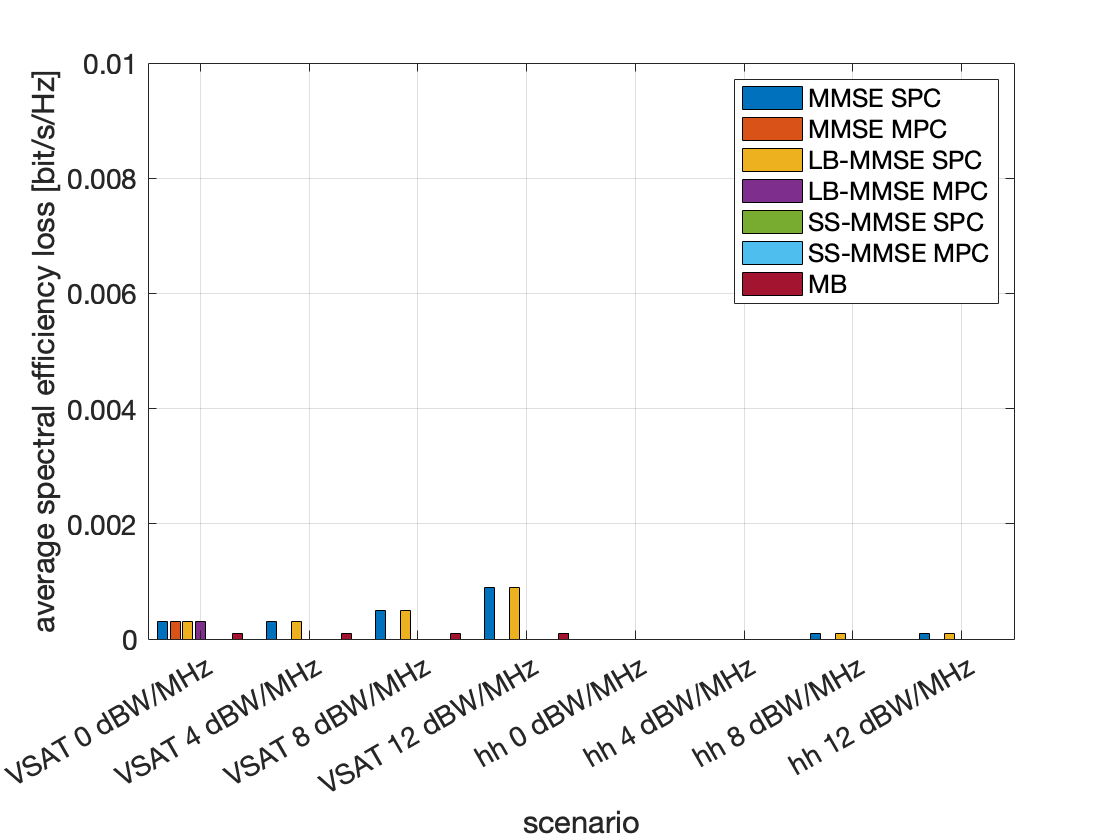}
        \label{fig:single_loss_rate_tlos}
    }
    \subfigure[$N_{node}=2$]
    {
        \includegraphics[width=0.48\textwidth]{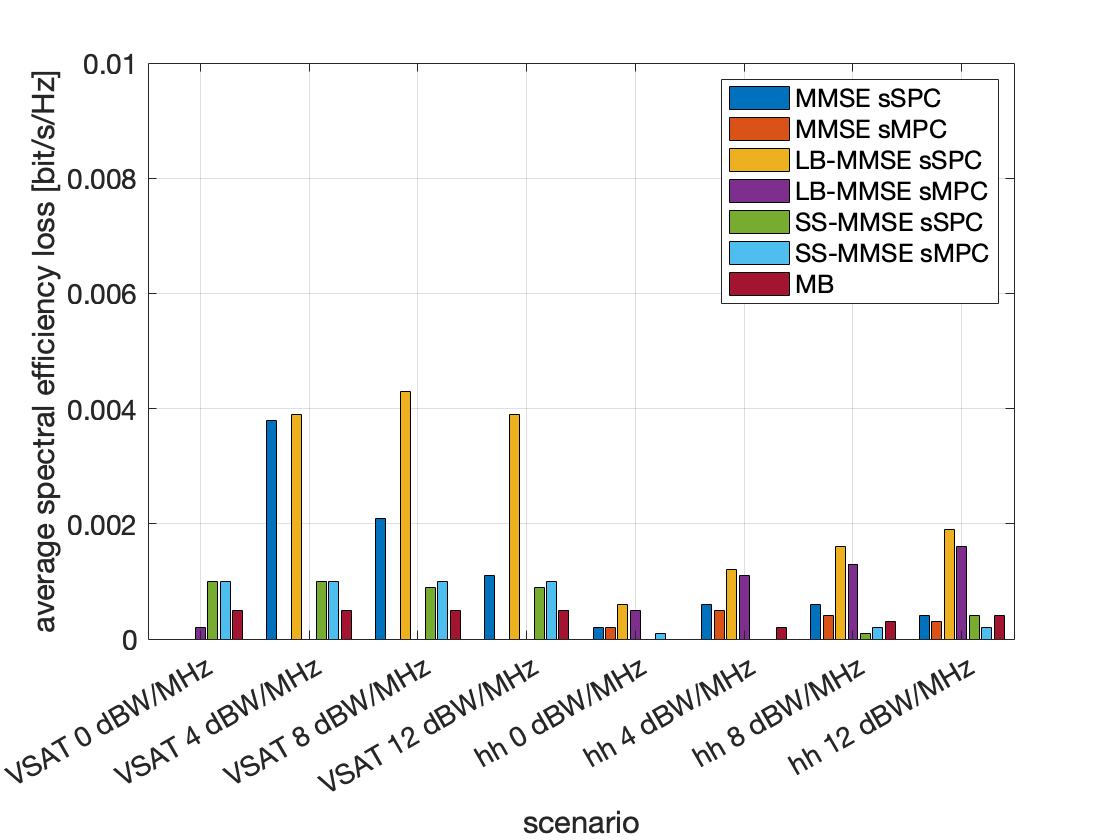}
        \label{fig:dual_loss_rate_tlos}
    }
    \caption{Loss in the average spectral efficiency in clear-sky conditions with $N_{node}=1$ (left) and $N_{node}=2$ (right).}
    \label{fig:tlos_loss}
\end{figure*}
\begin{figure*}
    \centering
    \subfigure[$N_{node}=1$]
    {
        \includegraphics[width=0.48\textwidth]{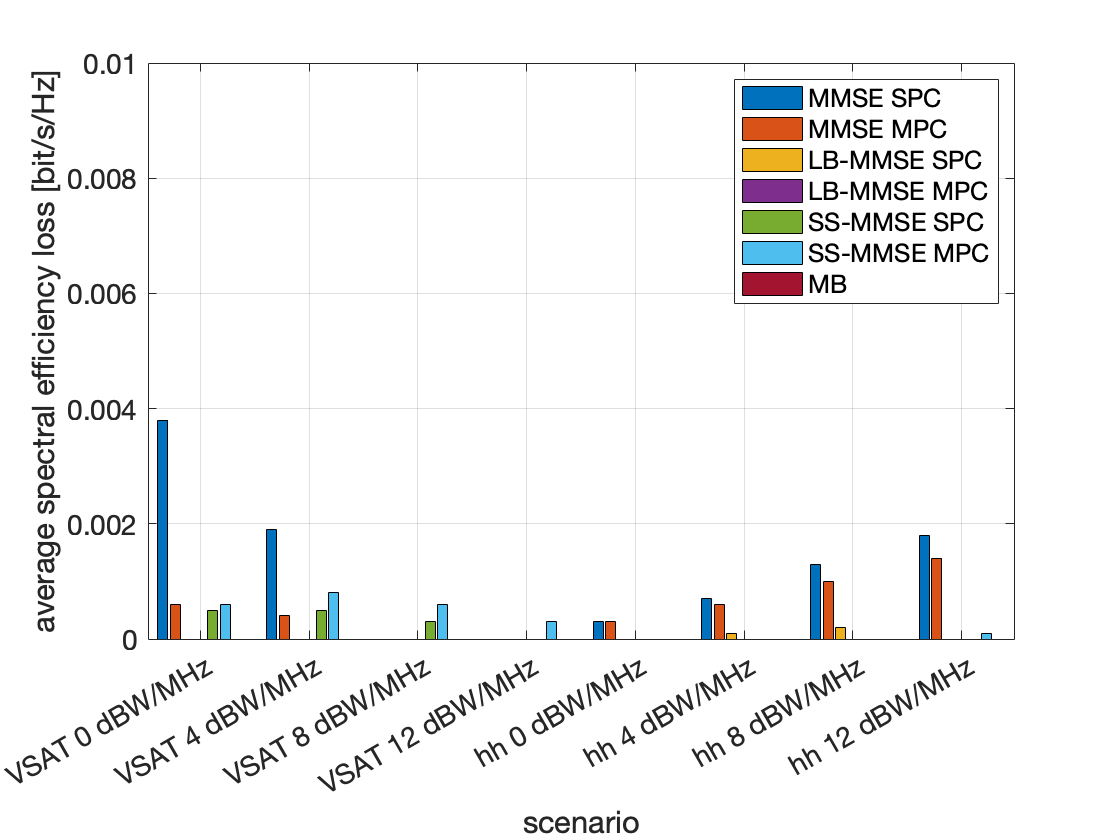}
        \label{fig:single_loss_rate_nlos}
    }
    \subfigure[$N_{node}=2$]
    {
        \includegraphics[width=0.48\textwidth]{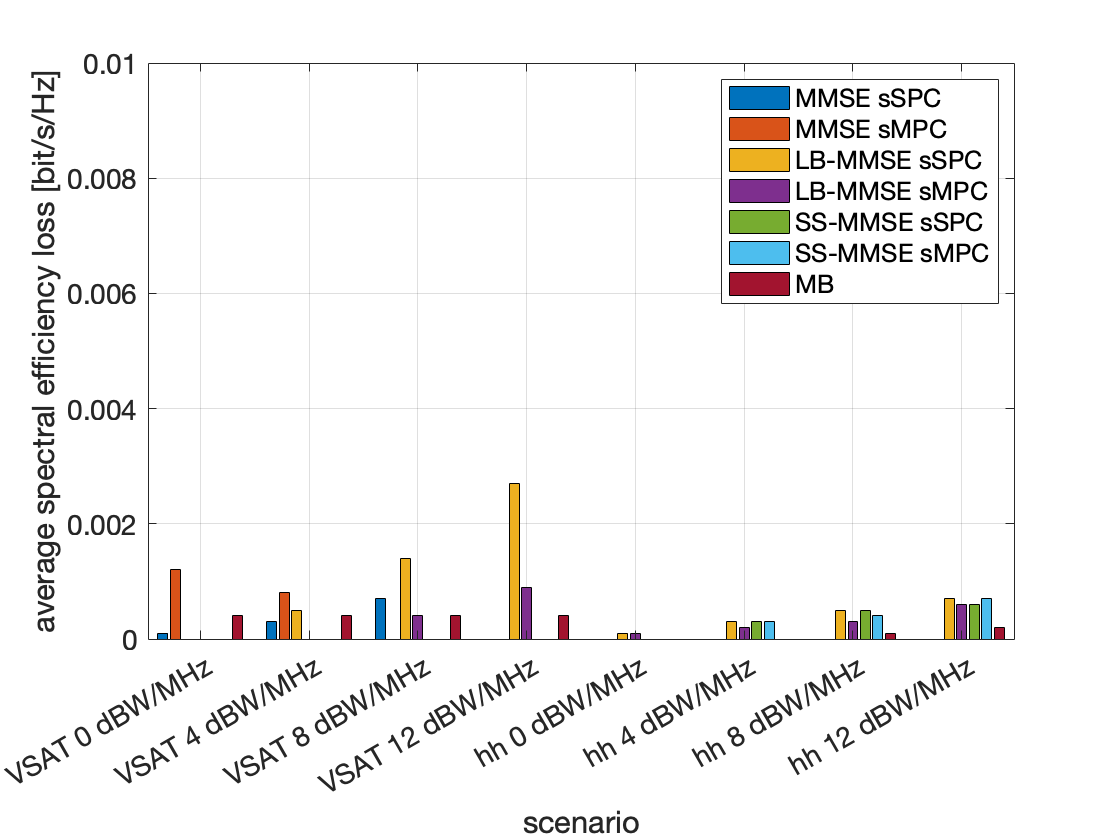}
        \label{fig:dual_loss_rate_nlos}
    }
    \caption{Loss in the average spectral efficiency in NLOS dense-urban conditions with $N_{node}=1$ (left) and $N_{node}=2$ (right).}
    \label{fig:nlos_loss}
\end{figure*}

Figures \ref{fig:nlos_rate} and \ref{fig:nlos_out} show the performance on the NLOS dense-urban channel. In general, we can observe that the performance is worse compared to clear-sky conditions in all cases, as expected. Moreover:
\begin{itemize}
	\item The LB-MMSE algorithm has a slightly worse performance compared to MMSE. Since this channel includes the additional losses, which are not taken into account by LB-MMSE, the misalignment between $\mathbf{W}^{(t_0)}$ and $\mathbf{H}^{(t_1)}$ is larger with this algorithm. However, considering all of the advantages that location-based solutions provide, extensively discussed in Section~\ref{sec:model_location}, this algorithm is indeed a viable and effective approach to implement CF-MIMO in NTN systems. The performance with SS-MMSE and MB is still much worse compared to MMSE and LB-MMSE, with a loss in the spectral efficiency up to $4$ bit/s/Hz with VSAT receivers.
	\item With respect to the normalisations, as in clear-sky conditions, (s)SPC performs better than (s)MPC, which however is the practically implementable solution.
	\item Comparing the centralised and federated scenarios, we can observe that the spectral efficiency is always better with a single node. In particular, with large transmission power and VSAT receivers, its gain can be as large as $2$ bit/s/Hz. However, observing the percentage of unserved users, it can be noticed that, with MMSE and LB-MMSE, the exploitation of a second node in the swarm is significantly beneficial with: i) low transmission power and VSAT; and ii) large transmission power and handheld. Thus, the exploitation of additional nodes is beneficial in harsh propagation conditions, since it introduces a gain in terms of path diversity. In dense-urban environments, the advantage of a second node in terms of outage can be: i) with low transmission power, as large as $6.1\%$ for MMSE-MPC with VSAT terminals; and ii) with large transmission power, as large as $5.7\%$ for MMSE-SPC with handheld terminals. In these conditions, despite the lower spectral efficiency, the system would be able to serve many more users with two nodes.
\end{itemize}

\subsection{Non-ideal information}
\label{sec:assessment_ni}
In this Section, we discuss the performance obtained when non-ideal information on the UEs' position and the radiation pattern model is considered for LB-MMSE and SS-MMSE. 
\paragraph{Positioning}
In this case, we assume that each user estimates its position with a uniformly distributed error of $\mathcal{U}\left[0,10\right)$ meters, in a random direction $\mathcal{U}\left[0,2\pi\right)$, with respect to its correct location. It shall be noticed that the maximum location error is significantly larger compared to the accuracy that current receivers equipped with GNSS can achieve. This is aimed at showing the significant robustness of the proposed LB-MMSE algorithm to a non-ideal estimation of the location.\\
Figures \ref{fig:tlos_loss} and \ref{fig:nlos_loss} show the loss in the average spectral efficiency in clear-sky and NLOS dense-urban conditions, respectively. At most, the performance is degraded by $0.04$ bit/s/Hz. This is in line with the previous statement related to the movement of the users (\emph{i.e.}, even when moving at $250$ km/h, the performance loss is in the order of $10^{-4}$ bit/s/Hz). In terms of the percentage of unserved users, the loss was observed to be negligible. Thus, LB-MMSE and SS-MMSE are significantly robust to positioning errors. 

\begin{figure*}
    \centering
    \subfigure[LB-MMSE]
    {
        \includegraphics[width=0.48\textwidth]{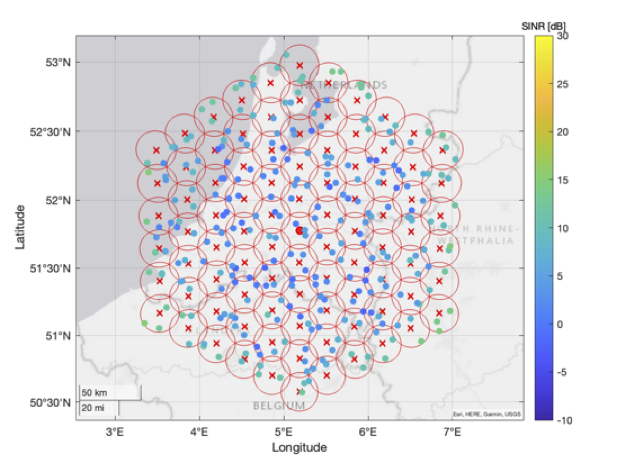}
        \label{fig:rp_3}
    }
    \subfigure[SS-MMSE]
    {
        \includegraphics[width=0.48\textwidth]{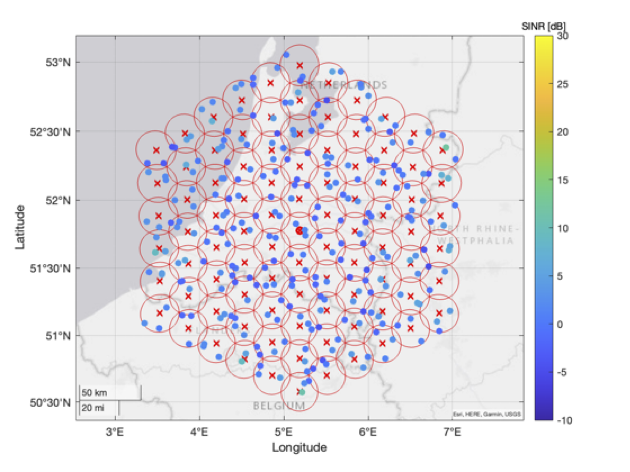}
        \label{fig:rp_4}
    }
    \caption{Geographical distribution of the SINR over $10$ time slots with LB-MMSE (left) and SS-MMSE (right) in NLOS dense-urban conditions with $\varepsilon_{rp}=0.05$ and a centralised node.}
    \label{fig:rp_b}
\end{figure*}

\begin{figure*}
    \centering
    \subfigure[$\varepsilon_{rp}=0$]
    {
        \includegraphics[width=0.48\textwidth]{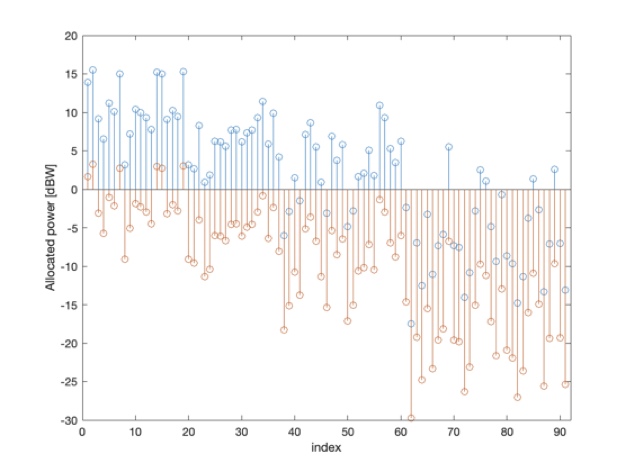}
        \label{fig:all_1}
    }
    \subfigure[$\varepsilon_{rp}=0.05$]
    {
        \includegraphics[width=0.48\textwidth]{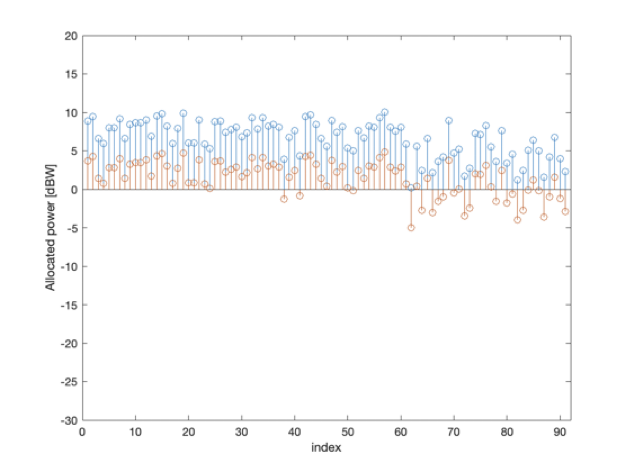}
        \label{fig:all_2}
    }
    \caption{Power allocation per served user with LB-MMSE with centralised MIMO.}
    \label{fig:all_a}
\end{figure*}

\begin{figure*}
    \centering
    \subfigure[$\varepsilon_{rp}=0$]
    {
        \includegraphics[width=0.48\textwidth]{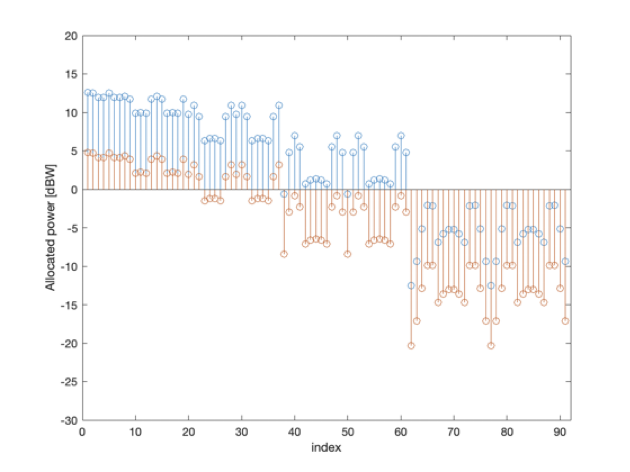}
        \label{fig:all_3}
    }
    \subfigure[$\varepsilon_{rp}=0.05$]
    {
        \includegraphics[width=0.48\textwidth]{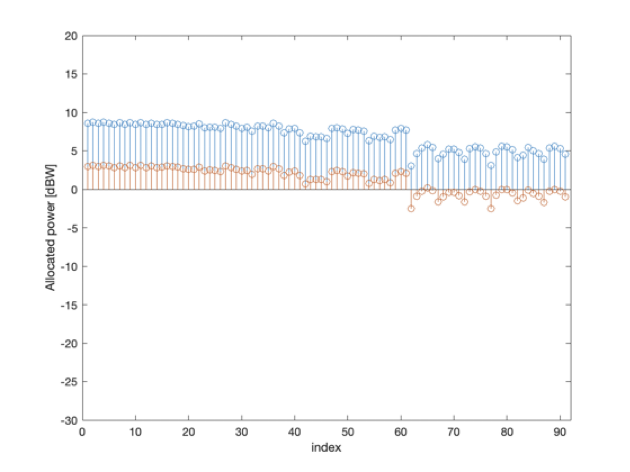}
        \label{fig:all_4}
    }
    \caption{Power allocation per served user (beam) with SS-MMSE with centralised MIMO.}
    \label{fig:all_b}
\end{figure*}

\paragraph{Radiation pattern model}
When computing the LB-MMSE and SS-MMSE channel coefficients based on the UEs' location, another source of error is the non-ideal representativeness of the radiation pattern model. To model this impairment, we focus on the on-board radiation pattern model. In particular, we assume that the estimated antenna radiation between the $n$-th element on-board the $s$-th satellite and the $i$-th user at the $t$-th time instant, $\widehat{g}_{i,n,s}^{(TX,t)}$, is given by:
\begin{equation}
	\widehat{g}_{i,n,s}^{(TX,t)} = g_{i,n,s}^{(TX,t)} + \Delta g_{i,n,s}^{(TX,t)}
\end{equation}
where $\Delta g_{i,n,s}^{(TX,t)} = \left|\Delta g_{i,n,s}^{(TX,t)}\right|e^{-\jmath \angle \Delta g_{i,n,s}^{(TX,t)}}$ is a r.v. modelling the error on the radiation pattern with the following amplitude and phase statistics: 
\begin{align}
	\left|\Delta g_{i,n,s}^{(TX,t)}\right|&\sim \mathcal{N}\left(0,{\left|g_{i,n,s}^{(TX,t)}\right|}^2\varepsilon_{rp}\right)\\
	\angle\Delta g_{i,n,s}^{(TX,t)}&\sim \mathcal{N}\left(0,{\left|\angle g_{i,n,s}^{(TX,t)}\right|}^2\varepsilon_{rp}\right)
\end{align}
The coefficient $\varepsilon_{rp}$ allows to adjust the variance of the amplitude and phase errors. In the following, we assume $\varepsilon_{rp}=0.05$, \emph{i.e.}, a $5\%$ error on the correct amplitude and phase of the radiation pattern. \\
\begin{figure*}
	\centering
	\subfigure[$N_{node}=1$]
	{
		\includegraphics[width=0.48\textwidth]{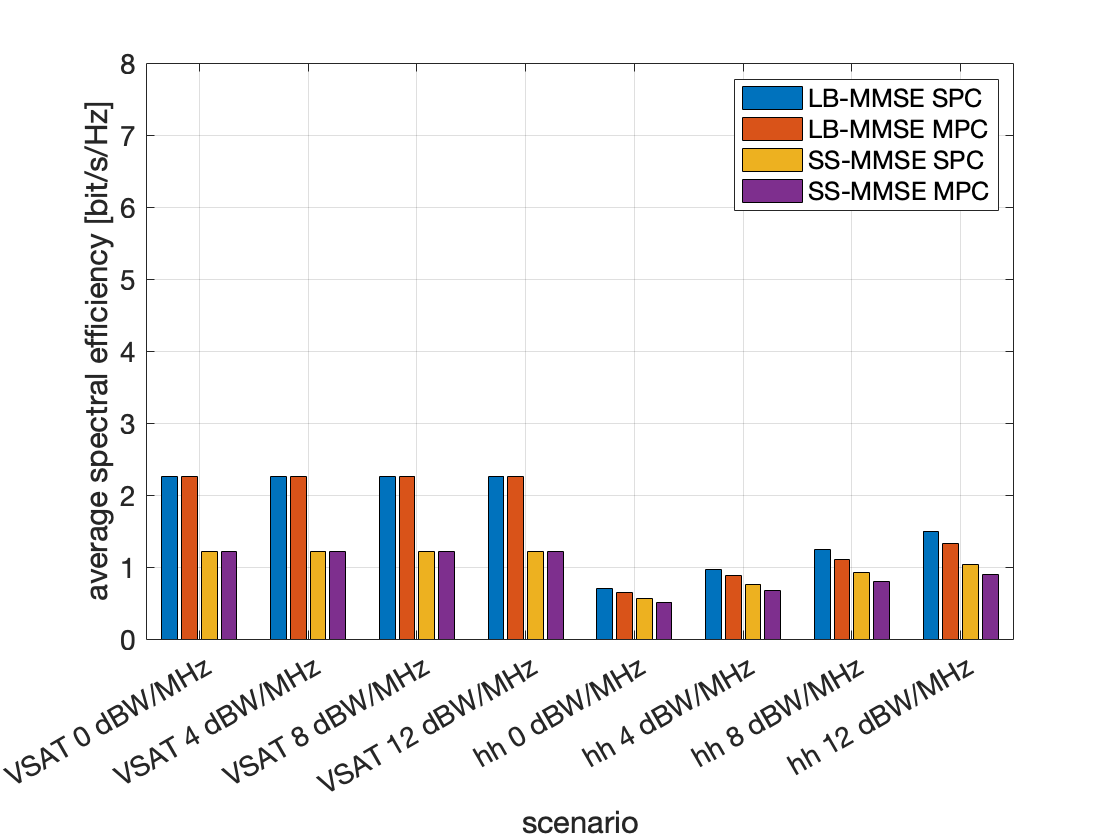}
		\label{fig:single_rate_rad_tlos}
	}
	\subfigure[$N_{node}=2$]
	{
		\includegraphics[width=0.48\textwidth]{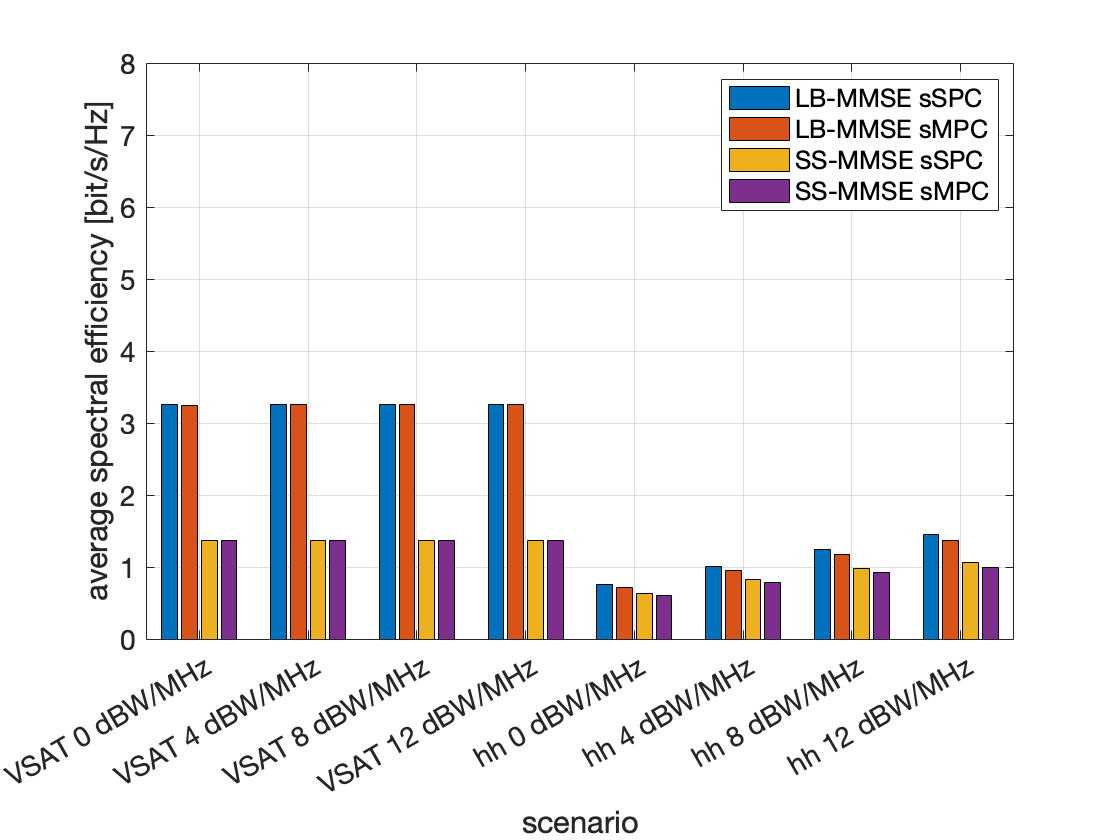}
		\label{fig:dual_rate_rad_tlos}
	}
	\caption{Average spectral efficiency in clear-sky with $N_{node}=1$ (left) and $N_{node}=2$ (right) and non-ideal radiation pattern.}
	\label{fig:tlos_rate_rad}
\end{figure*}
\begin{figure*}
	\centering
	\subfigure[$N_{node}=1$]
	{
		\includegraphics[width=0.48\textwidth]{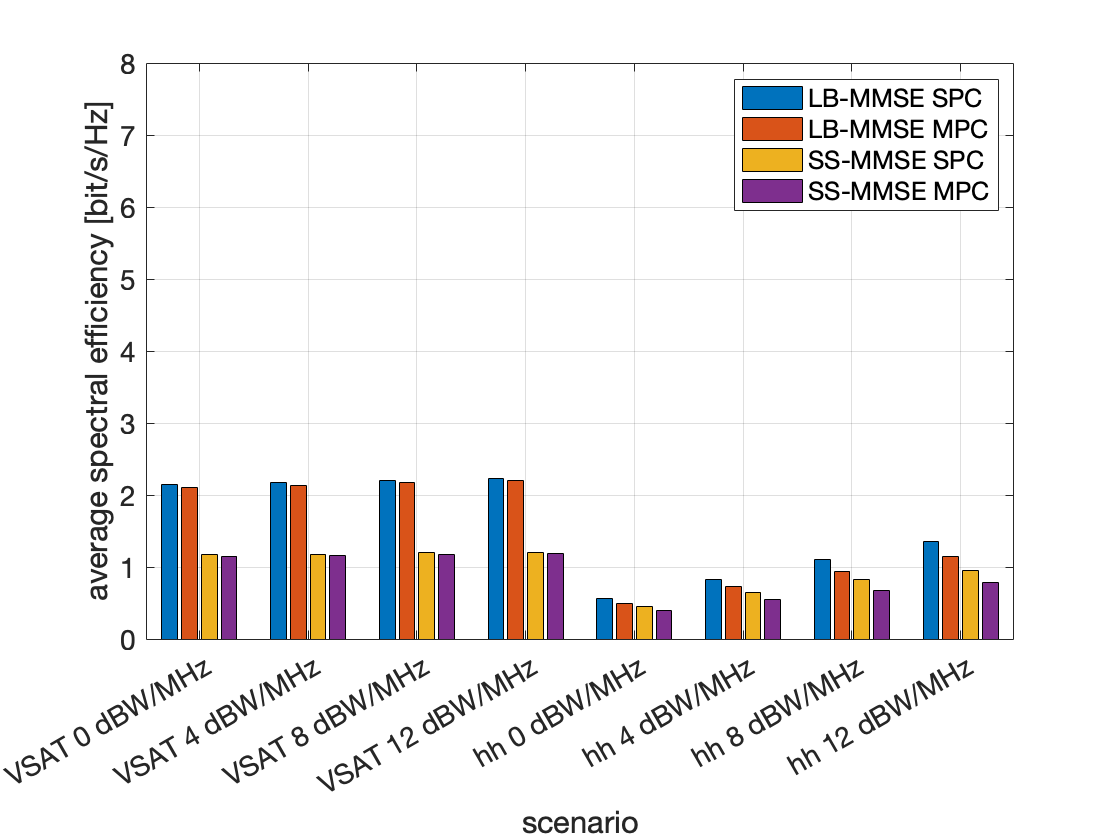}
		\label{fig:single_rate_rad_nlos}
	}
	\subfigure[$N_{node}=2$]
	{
		\includegraphics[width=0.48\textwidth]{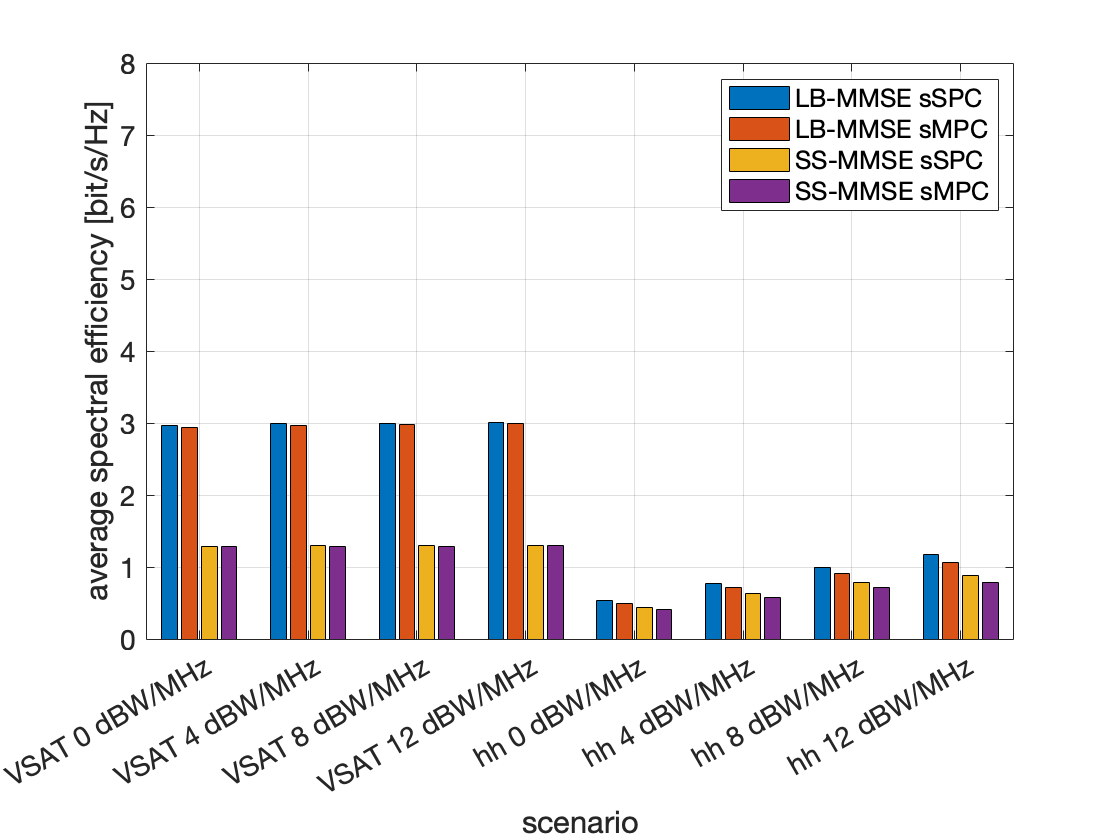}
		\label{fig:dual_rate_rad_nlos}
	}
	\caption{Average spectral efficiency in NLOS dense-urban conditions with $N_{node}=1$ (left) and $N_{node}=2$ (right) and non-ideal radiation pattern.}
	\label{fig:nlos_rate_rad}
\end{figure*}
\begin{figure*}
    \centering
    \subfigure[$N_{node}=1$]
    {
        \includegraphics[width=0.48\textwidth]{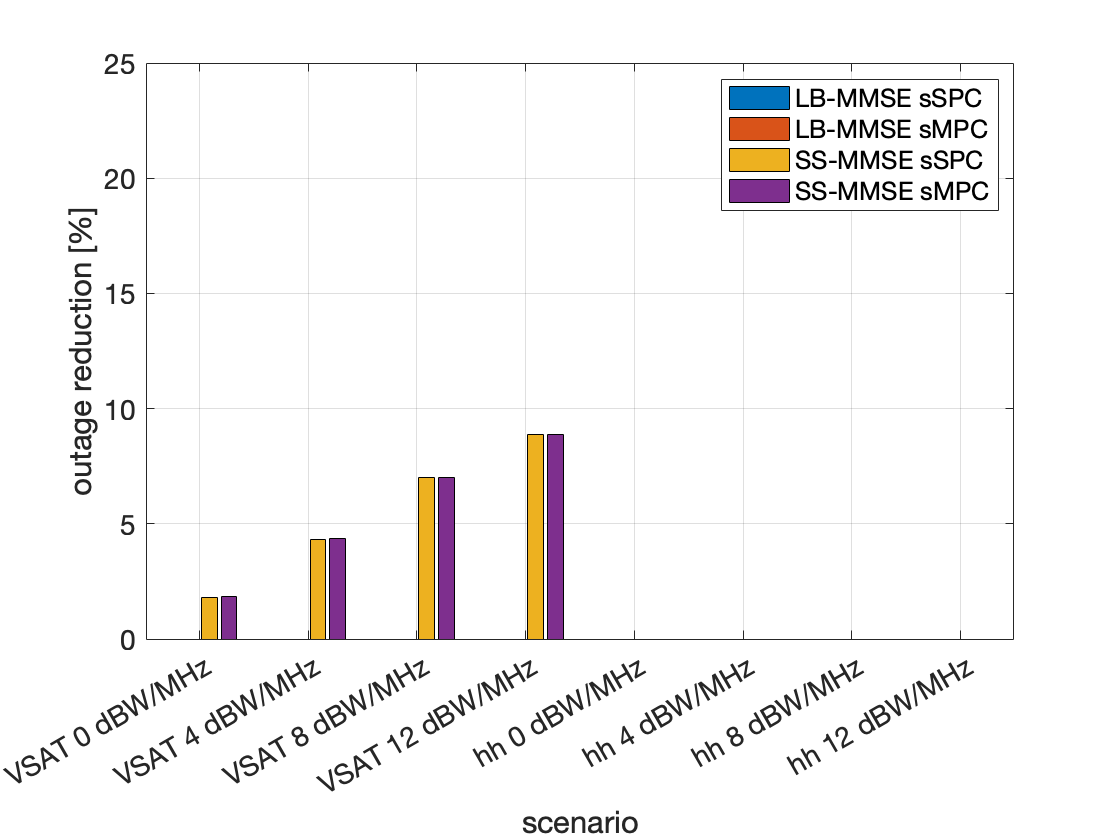}
        \label{fig:outage_single_tlos}
    }
    \subfigure[$N_{node}=2$]
    {
        \includegraphics[width=0.48\textwidth]{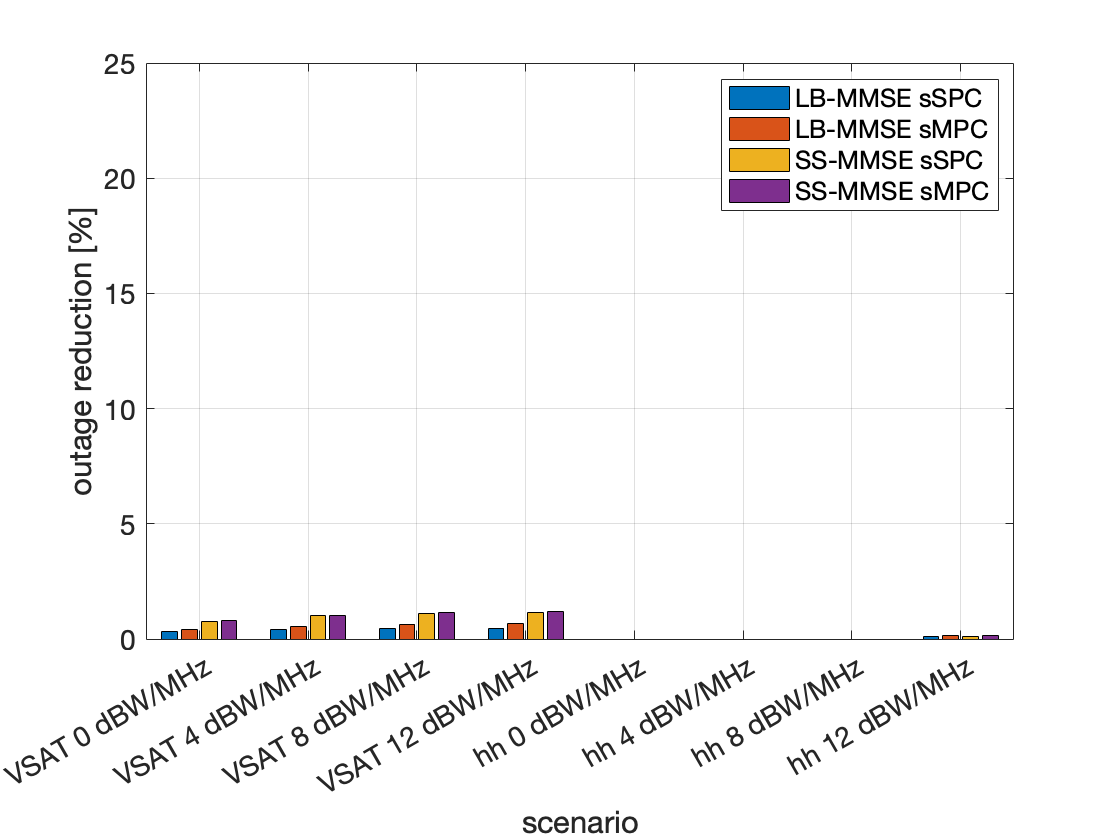}
        \label{fig:outage_dual_tlos}
    }
    \caption{Reduction in the percentage of unserved users in clear-sky with $N_{node}=1$ (left) and $N_{node}=2$ (right) and $\varepsilon_{rp}=0.05$.}
    \label{fig:tlos_out_rad}
\end{figure*}
\begin{figure*}
    \centering
    \subfigure[$N_{node}=1$]
    {
        \includegraphics[width=0.48\textwidth]{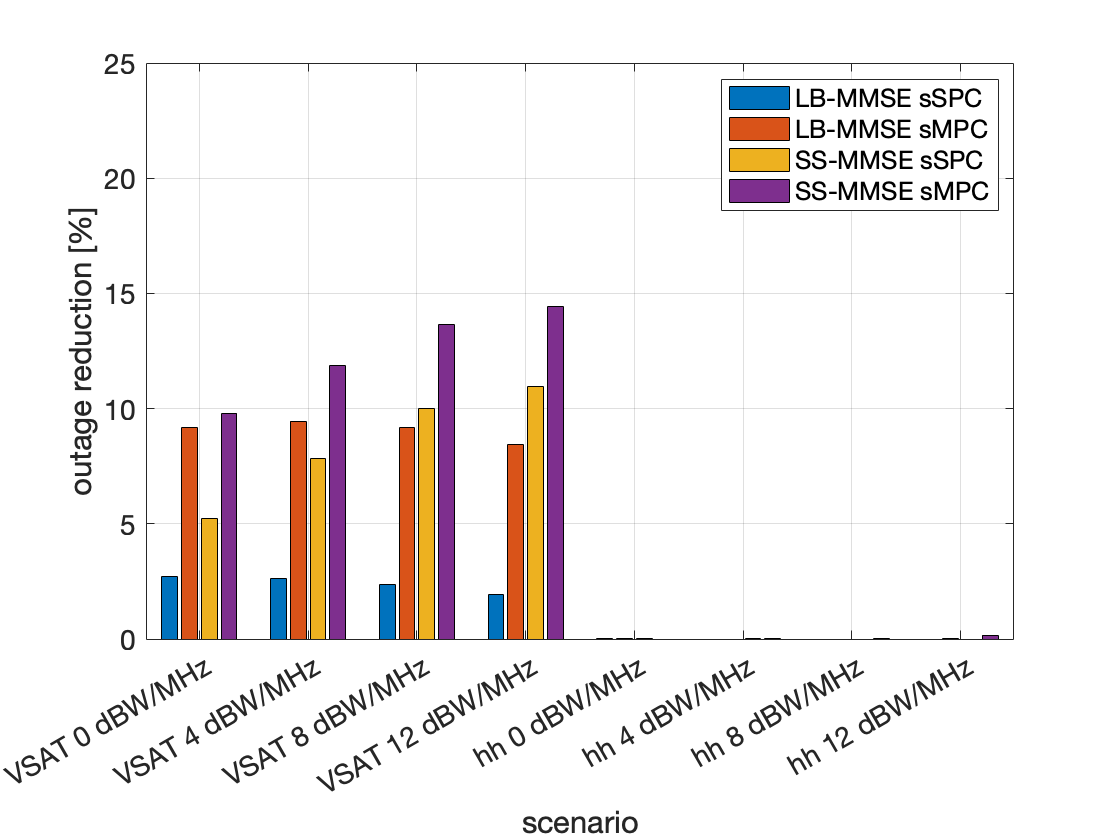}
        \label{fig:outage_single_nlos}
    }
    \subfigure[$N_{node}=2$]
    {
        \includegraphics[width=0.48\textwidth]{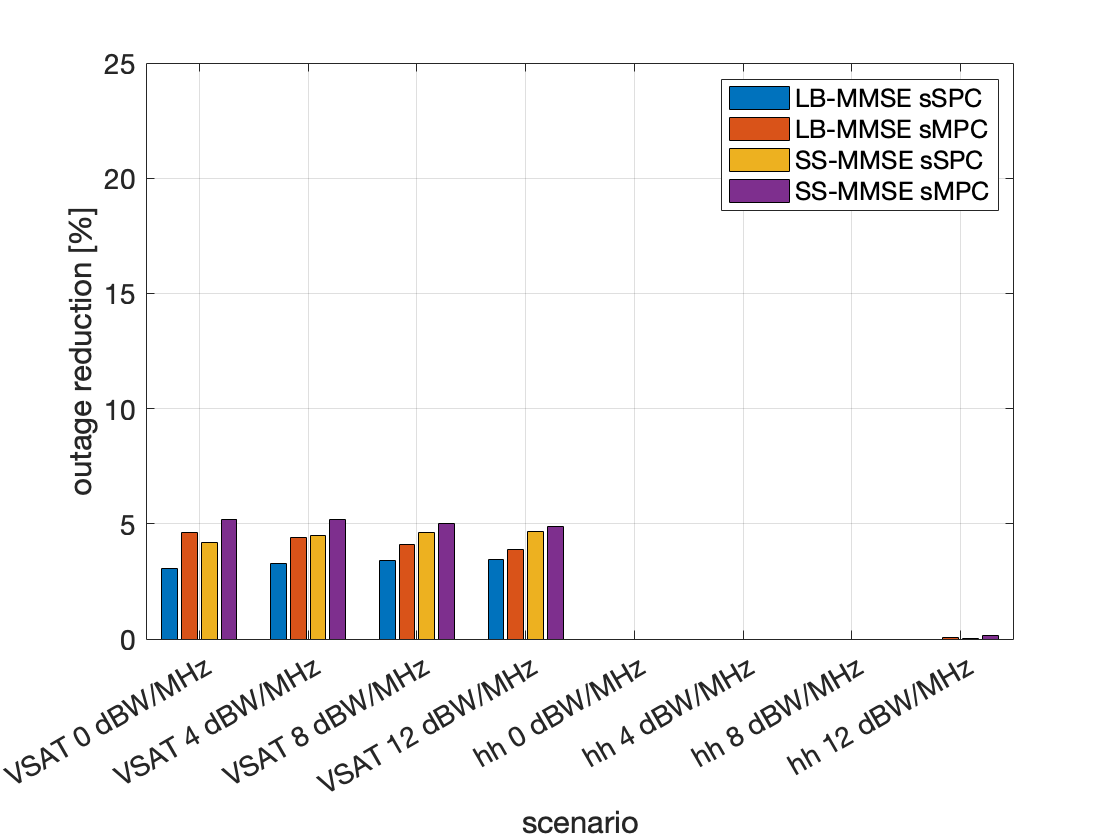}
        \label{fig:outage_dual_nlos}
    }
    \caption{Reduction in the percentage of unserved users in NLOS dense-urban conditions with $N_{node}=1$ (left) and $N_{node}=2$ (right) and $\varepsilon_{rp}=0.05$.}
    \label{fig:nlos_out_rad}
\end{figure*}
Figures \ref{fig:tlos_rate_rad} and \ref{fig:nlos_rate_rad} show the average spectral efficiency for the LB-MMSE and SS-MMSE algorithms and $\varepsilon_{rp}=0.05$. It can be noticed that both algorithms are particularly impacted by a non-ideal knowledge. The loss with VSATs in the centralised scenario can be as large as $4.5$ bit/s/Hz and $3.5$ bit/s/Hz in clear-sky and NLOS dense-urban conditions, respectively; with federated solutions and two nodes, the performance is less degraded, since the system benefits from the transmission from distributed sources, and the loss is in the order of $1-1.5$ bit/s/Hz. With handheld terminals, the performance with two nodes is still better, but the advantage is more limited compared to the centralised case. \\
In terms of unserved users, some interesting behaviours arise. In particular, with VSAT terminals the performance with an error on the radiation pattern model is actually improved in terms of outage, while this phenomenon is absent for handheld terminals. Figures \ref{fig:tlos_out_rad} and \ref{fig:nlos_out_rad} show the outage reduction representing such gain; it can be noticed that in the centralised scenario this phenomenon is more relevant. To understand the motivation for this behaviour, let us focus on the centralised case in NLOS dense-urban conditions, where it is more evident. In particular, Figures \ref{fig:rp_a} and \ref{fig:rp_b} show the geographical distribution of the SINR with $\varepsilon_{rp}=0$ (ideal), and $\varepsilon_{rp}=0.05$, respectively. It can be noticed that:
\begin{figure*}
    \centering
    \subfigure[LB-MMSE]
    {
        \includegraphics[width=0.48\textwidth]{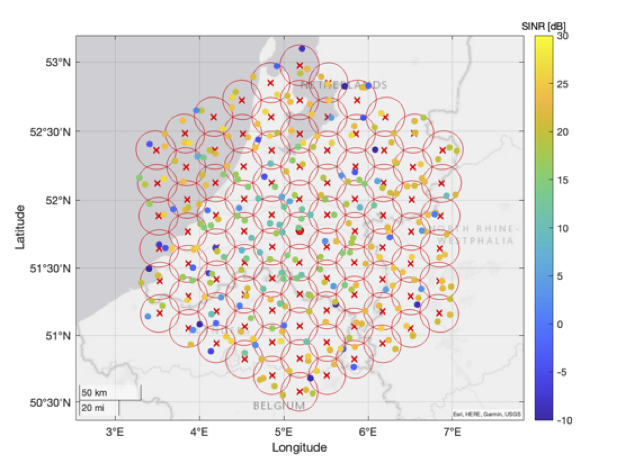}
        \label{fig:rp_1}
    }
    \subfigure[SS-MMSE]
    {
        \includegraphics[width=0.48\textwidth]{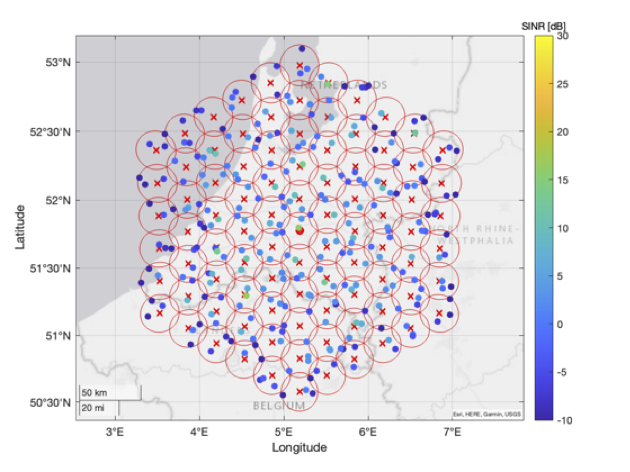}
        \label{fig:rp_2}
    }
    \caption{Geographical distribution of the SINR over $10$ time slots with LB-MMSE (left) and SS-MMSE (right) in NLOS dense-urban conditions with $\varepsilon_{rp}=0$ and a centralised node.}
    \label{fig:rp_a}
\end{figure*}
\begin{itemize}
	\item With LB-MMSE and $\varepsilon_{rp}=0$, the users in the outer part of the coverage are those in the best channel conditions; the uniformly distributed users that are experiencing a bad SINR are those in NLOS conditions, which include the clutter loss and a harsh standard deviation of the shadow fading. The best behaviour at the coverage edge is due to the combination of the reduced interference in the external areas of the coverage region and the nature of LB-MMSE, which is a user-centric technique in which the CSI vectors are identified specifically per user without approximations, compared to SS-MMSE. When $\varepsilon_{rp}=0.05$, the performance at the edge of the coverage is still better compared to the inner area, but in general the SINR is significantly lower and much more uniformly distributed, leading to a loss in the spectral efficiency and a gain in the outage, since all users are now above the SINR threshold. The error on the radiation pattern is making the CSI vectors more uniform from the beamformer perspective and all users are served with a similar SINR (as also demonstrated by the constant performance in Figures \ref{fig:tlos_out_rad} and \ref{fig:nlos_out_rad} for increasing values of the power density). This is further substantiated in Figure~\ref{fig:all_a}, which shows the power allocated to the users in a time slot with LB-MMSE: i) in ideal conditions, the users at the coverage edge are allocated less power because they experience less interference and also have large antenna gains; ii) when the error is introduced, the power tends to be uniformly distributed across the beams, since the users' CSI vectors tend to be more and more similar.
	\item SS-MMSE is a beam-based algorithm, as each user is approximated with the closest beam center. Thus, the performance is driven by the distance from the associated beam center: the larger this distance, the larger the approximation and, consequently, the worst the performance. In addition to this aspect, also with SS-MMSE the external beams are allocated a lower power; however, differently from LB-MMSE, the reduced interference in the external tiers is not sufficient to cope with both the lower allocated power and the approximation in the channel coefficients. Thus, the external tiers experience a worse performance compared to the inner ones. When $\varepsilon_{rp}=0.05$, the same phenomenon described above for LB-MMSE arises. The CSI vectors tend to become more similar from the beamformer perspective and a uniform performance is achieved across the coverage area. This is substantiated in Figure~\ref{fig:all_b}, with a very limited variability in the power allocations to the beams when $\varepsilon_{rp}=0.05$. As observed above, the consequence is that the spectral efficiency is worse compared to the ideal scenario, while the outage is improved since users at the coverage edge that were below the SINR threshold now experience an SINR close to that of all users.
\end{itemize}
The above phenomenon is still present for SS-MMSE in clear-sky, while for the LB-MMSE it is not. In clear-sky conditions, no user is experiencing the shadow fading or clutter loss; thus, with an algorithm in which the actual locations are used (and not an approximation), no user is in outage at the coverage edge and only the loss in the spectral efficiency is present. Similar considerations and observations can be made for the federated scenario.
\begin{figure*}
    \centering
    \subfigure[$N_{node}=1$]
    {
        \includegraphics[width=0.48\textwidth]{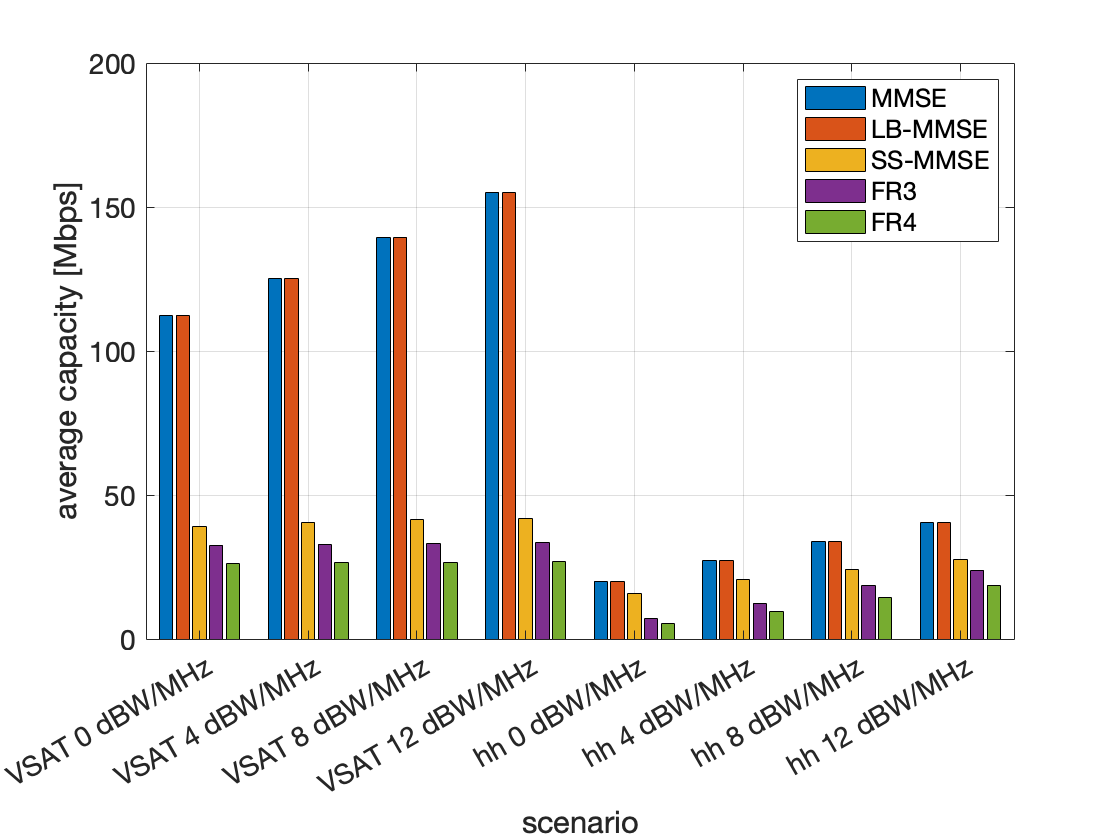}
        \label{fig:capacity_1}
    }
    \subfigure[$N_{node}=2$]
    {
        \includegraphics[width=0.48\textwidth]{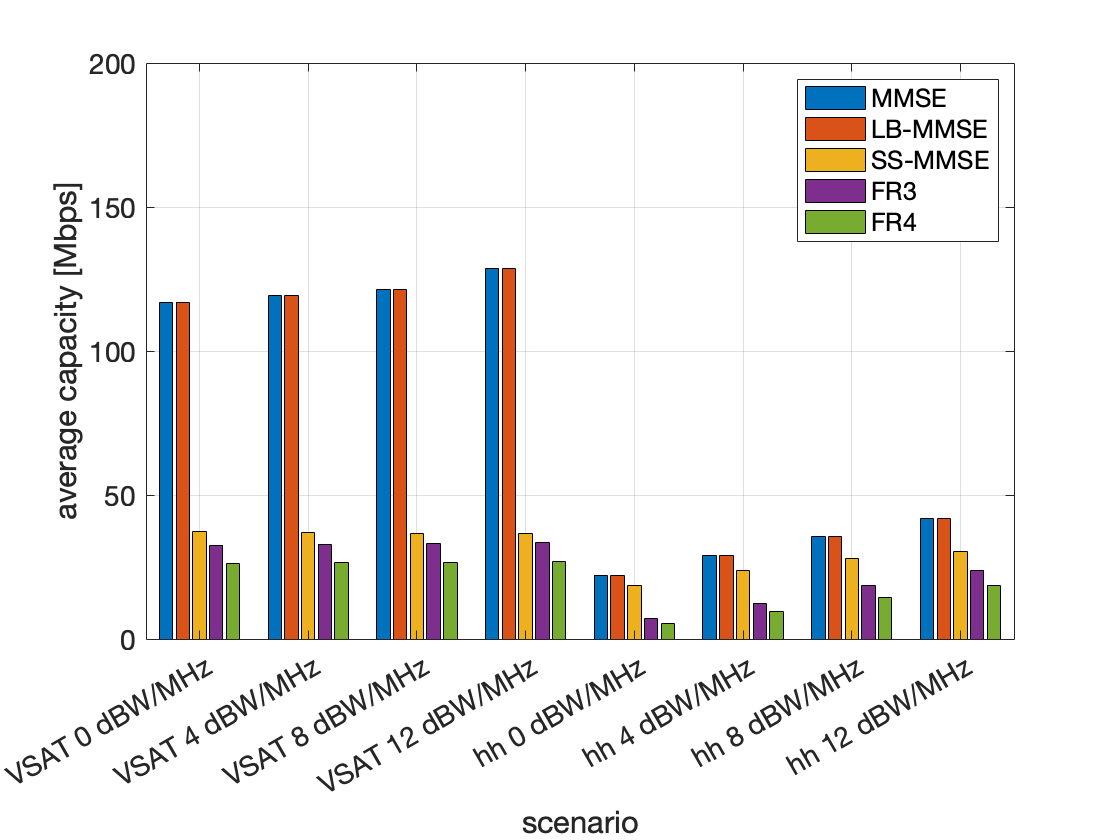}
        \label{fig:capacity_2}
    }
    \caption{Average capacity in clear-sky with $N_{node}=1$ (left) and $N_{node}=2$ (right).}
    \label{fig:capacity_a}
\end{figure*}
\begin{figure*}
    \centering
    \subfigure[$N_{node}=1$]
    {
        \includegraphics[width=0.48\textwidth]{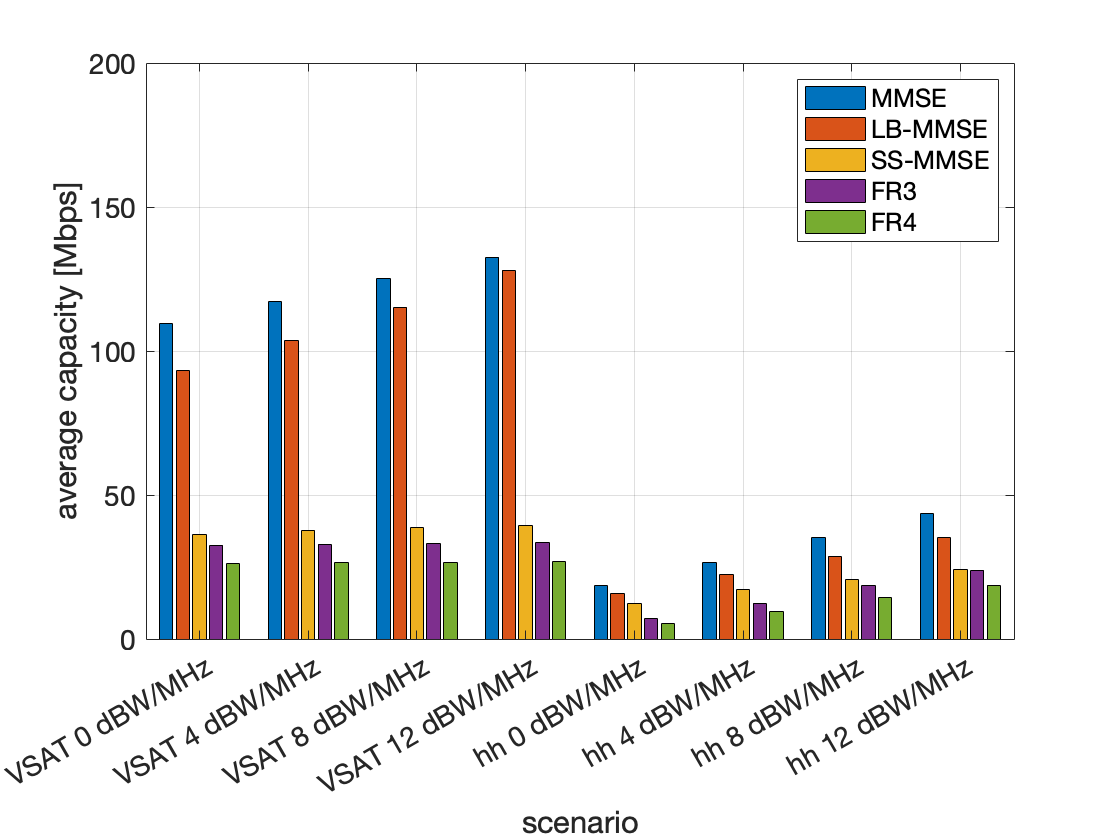}
        \label{fig:capacity_3}
    }
    \subfigure[$N_{node}=2$]
    {
        \includegraphics[width=0.48\textwidth]{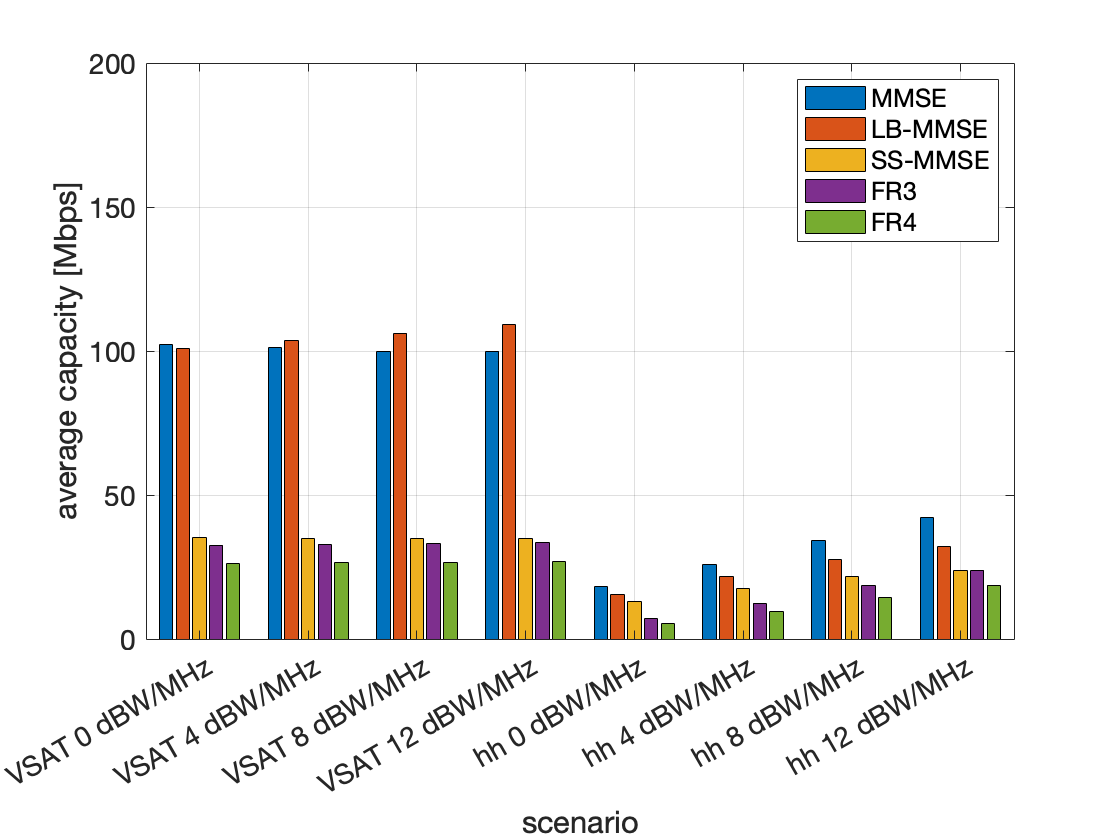}
        \label{fig:capacity_4}
    }
    \caption{Average capacity in NLOS dense-urban with $N_{node}=1$ (left) and $N_{node}=2$ (right).}
    \label{fig:capacity_b}
\end{figure*}
\subsection{Frequency reuse}
To conclude the extensive numerical assessment of the considered CF-MIMO techniques, we now discuss the comparison with legacy frequency reuse schemes with 3 and 4 colours (FR3 and FR4) implemented with a single node. To this aim, we consider the (s)MPC normalisation only. Figures \ref{fig:capacity_a} and \ref{fig:capacity_b} show the average capacity in clear-sky and NLOS dense-urban conditions, respectively. It can be noticed that the advantage of using a FFR scheme is significant: i) for VSAT receivers, the gain is in the order of $100$ Mbps in NLOS dense-urban conditions and even larger in clear-sky; ii) with handheld terminals, the gain is in the order of $15-20$ Mbps and $20-35$ Mbps for clear-sky and NLOS dense-urban environments, respectively. In terms of outage:
\begin{figure*}
    \centering
    \subfigure[$N_{node}=1$]
    {
        \includegraphics[width=0.48\textwidth]{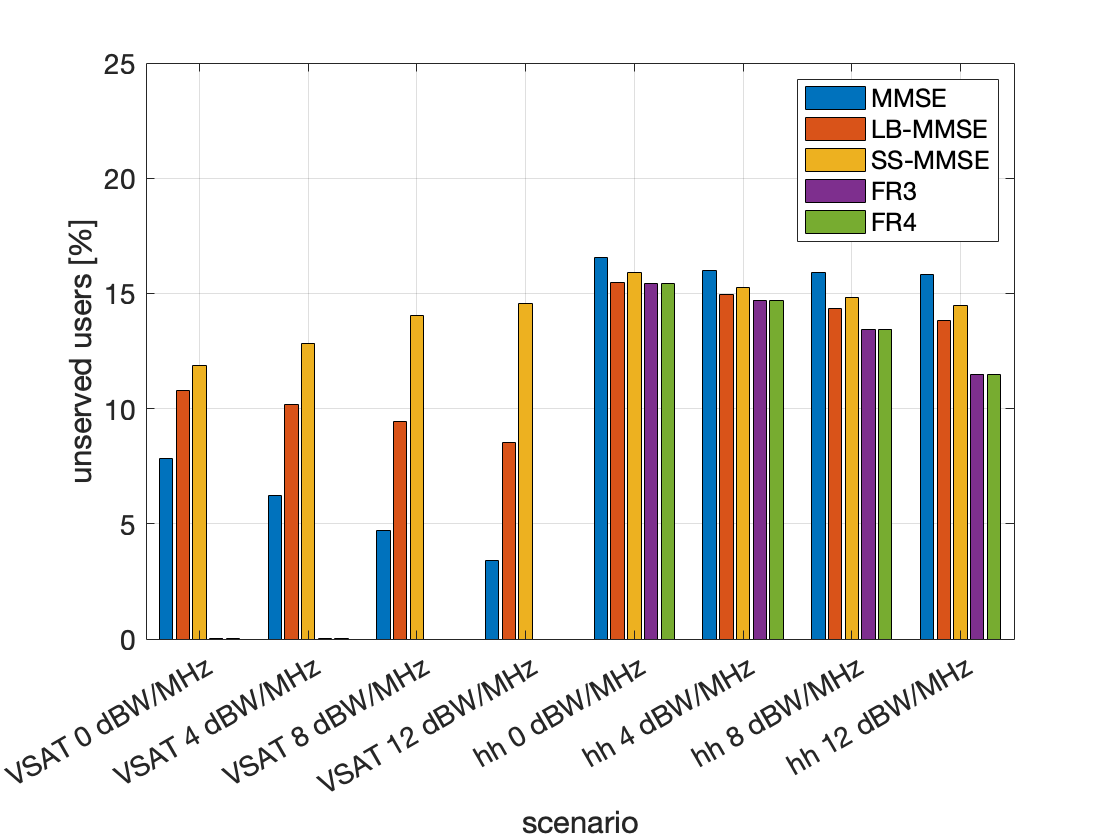}
        \label{fig:outage_3}
    }
    \subfigure[$N_{node}=2$]
    {
        \includegraphics[width=0.48\textwidth]{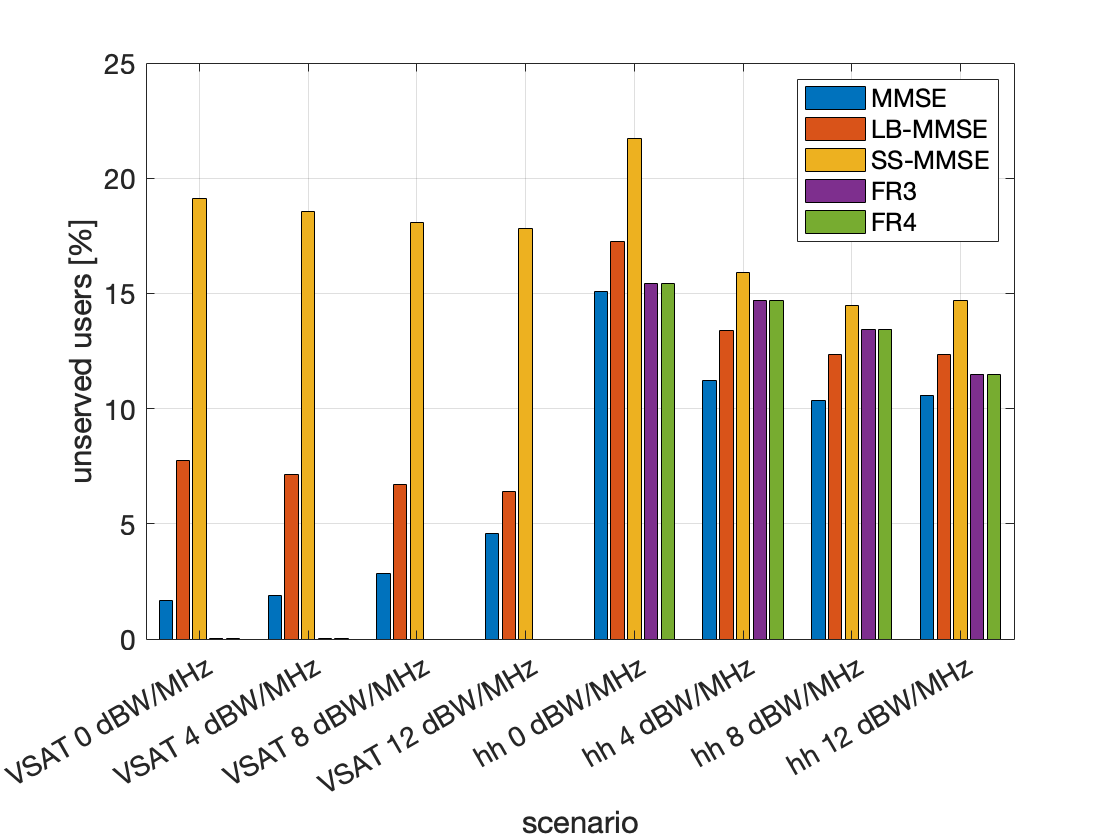}
        \label{fig:outage_4}
    }
    \caption{Average capacity in NLOS dense-urban with $N_{node}=1$ (left) and $N_{node}=2$ (right).}
    \label{fig:outage_a}
\end{figure*}
\begin{itemize}
	\item It was observed that no unserved users are present with 3 and 4 colours in clear-sky conditions; the outage performance of CF-MIMO on this channel (Figure~\ref{fig:tlos_out}) showed that for the (s)MPC normalisation a significant amount of unserved users was present only with two nodes and handheld terminals. Thus, for VSATs and handheld terminals in the centralised scenario the advantage of FFR with CF-MIMO techniques is significant.
	\item Figure~\ref{fig:outage_a} shows the unserved users in NLOS dense-urban conditions. As for VSATs, it can be observed that: i) with large transmission power, the MMSE algorithm shows a limited outage and, thus, the advantage of CF-MIMO is clear; ii) with low transmission power, the outage is a bit larger, but it might still be acceptable considering the gain in terms of capacity. As for handheld terminals, there is a significant amount of unserved users also with FR3 and FR4: i) in the centralised scenario, the outage is slightly below that with all CF-MIMO solutions, but considering the advantage in terms of capacity, the latter shall be selected; and ii) in the federated scenario, the path diversity introduced by the additional nodes provides an advantage over FR3 and FR4 also in terms of outage.
\end{itemize}
In conclusion, the capacity is significantly improved with CF-MIMO and FFR. However, by also taking into account the percentage of unserved users, there are some scenarios in which legacy systems might still be better, \emph{e.g.}, with VSAT terminals and low transmission power. In all other cases, the CF-MIMO approach is by far the best system-level solution.

\subsection{Recommendations}
\label{sec:recommendations}
Focusing on the (s)MPC normalisation, which is the feasible solution avoiding non-linear effects in the HPAs and limiting the power per node, the following  considerations hold:
\begin{itemize}
	\item The proposed LB-MMSE provides a performance identical to MMSE in clear-sky. The loss when taking into account additional stochastic losses, which cannot be \emph{a priori} estimated, is negligible even in harsh conditions, such as the considered NLOS dense-urban case.
	\item A federated Cell-Free solution shall be selected with VSAT terminals and lower transmission power (\emph{i.e.}, up to $0-2$ dBW/MHz). In this case, the achievable spectral efficiency is larger and the increase in the outage is negligible ($0.3-2.1\%$) in clear-sky conditions, while it is significantly lower in NLOS dense-urban conditions (up to $6.1\%$). With larger transmission power, a centralised solution shall be preferred.
	\item With handheld terminals, centralised beamforming is the best option based on the outage probability (a $0.1-0.2$ bit/s/Hz increase in the spectral efficiency is not enough to justify a $25\%$ outage) in clear/sky conditions, while in NLOS dense-urban environments federated solutions with multiple nodes provide an advantage in both the spectral efficiency ($0.02-0.2$ bit/s/Hz) and outage (up to $5.7\%$), thanks to the spatial diversity at the transmitter.
	\item Compared to FR3 and FR4 schemes, the capacity is significantly better with CF-MIMO solutions. When the outage is taken into account, CF-MIMO shall still be selected in the vast majority of the scenarios, with some limited exceptions (\emph{e.g.}, VSAT receivers and low transmission power).
\end{itemize}
Finally, when considering non-ideal information at the transmitter, it was observed that the proposed LB-MMSE algorithm: i) is particularly robust to positioning errors; ii) is significantly impacted by errors on the radiation pattern model. As for the latter, it is worthwhile highlighting that in real deployment scenarios the variance of the amplitude and phase errors on the antenna pattern coefficients is expected to be much lower than $5\%$ as assumed in the above analyses.

\section{Conclusions}
\label{sec:conclusions}
In this paper, we provided a detailed discussion on the design choices allowing the implementation of both federated and centralised Cell-Free or beam-based MIMO solutions in NTN NGSO constellations; both regenerative, with functional split options, and transparent payloads were addressed, as well as OGBF and OBBF solutions. A detailed description of the architecture options, with the related challenges and benefits, has been reported. Then, we designed: i) a novel location-based CF-MIMO algorithm for NTN NGSO constellations which is completely user-centric, \emph{i.e.}, tailored to the actual users' locations.; and ii) novel power normalisation approaches for federated MIMO algorithms that can be applied to swarms of NGSO nodes. The outcomes of the extensive numerical results showed that: i) the proposed LB-MMSE algorithm provides a performance close to the CSI-based MMSE, but with a significantly reduced overhead and complexity at the terminal side; ii) federated CF-MIMO solutions over NGSO swarms provide benefits with VSAT terminals and low transmission power, and with handheld terminals in NLOS dense-urban conditions. The performance was assessed also considering 3 and 4 colours frequency reuse schemes; it was shown that, apart from VSAT terminals and low transmission power, the CF-MIMO paradigm provides a significant performance improvement. Finally, the robustness of the considered location-based algorithms was assessed with non-ideal location estimation and non-ideal knowledge of the actual radiation of the on-board UPA elements, with respect to their mathematical model. It was shown that the former does not pose issues, while the latter is more critical. Future developments of this work include the analysis of the tight synchronisation among the swarm nodes for federated CF-MIMO and the performance assessment with variable UPA configurations and more than two nodes per swarm.

\bibliographystyle{IEEEtran}
\bibliography{2023_DYNASAT_ACCESS}

\begin{IEEEbiography}[{\includegraphics[width=1in,height=1.25in,clip,keepaspectratio]{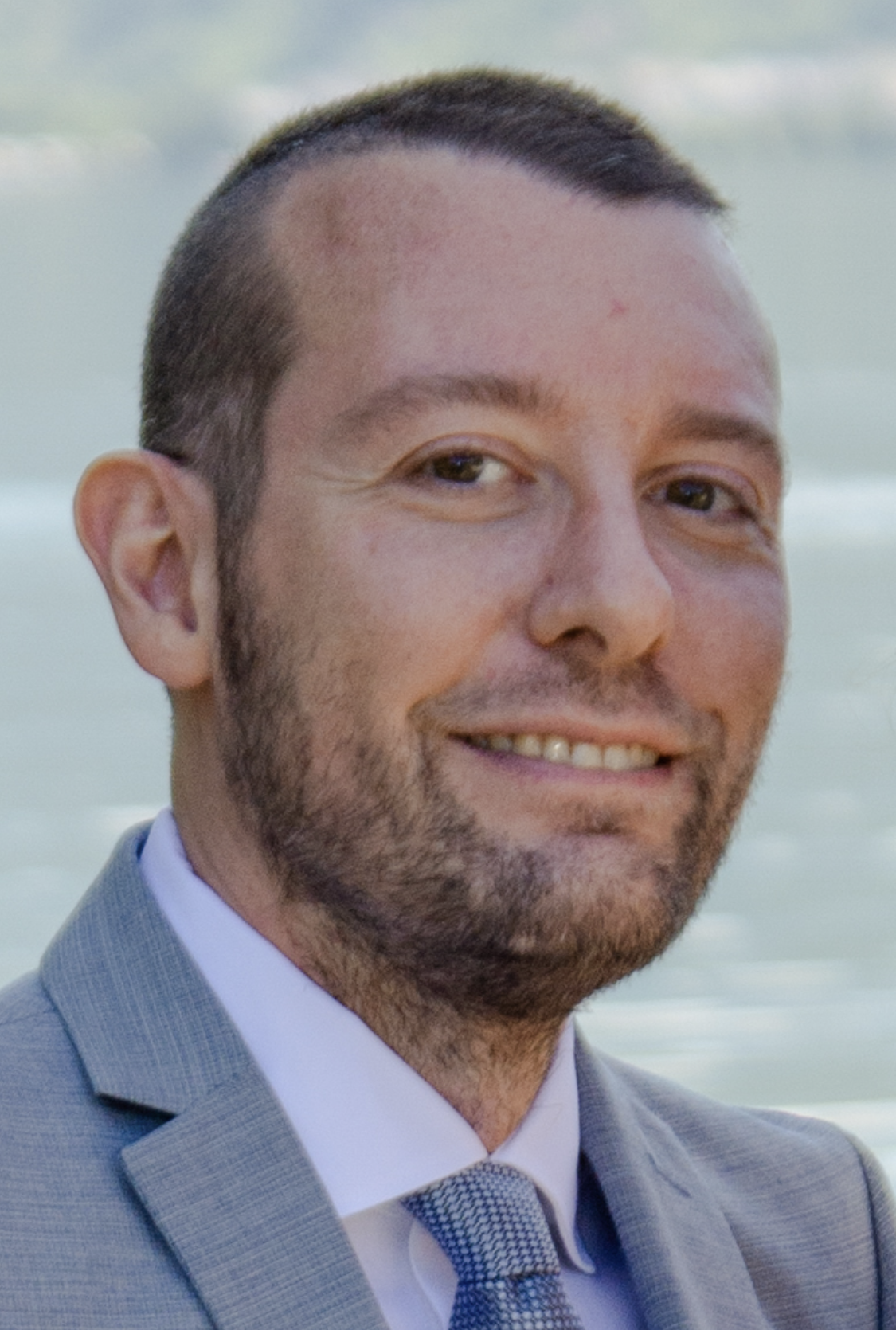}}]
	{Alessandro Guidotti} (Member, IEEE) received the master degree (magna cum laude) in telecommunications engineering and the Ph.D. degree in electronics, computer science, and telecommunications from the University of Bologna, Italy, in 2008 and 2012, respectively. From 2009 to 2011, he was a representative for the Italian Administration within CEPT SE43. In 2011 and 2012, he was a Visiting Researcher at SUPELEC, Paris, France. From 2014 to 2021, hew was a Research Associate with the Department of Electrical, Electronic, and Information Engineering ``Guglielmo Marconi,'' University of Bologna. From 2021, he is a Researcher with the Consorzio Inter-Universitario delle Telecomunicazioni (CNIT), located at the Research Unit of the University of Bologna. He is active in national and international research projects on wireless and satellite communication systems in several European Space Agency and European Commission funded projects. He is a member of the Editorial Board as Review Editor of the Aerial and Space Networks journal for Frontiers in Space Technologies. He has been serving as TPC and Publication Co-Chair at the ASMS/SPSC Conference since 2018. His research interests include wireless communication systems, spectrum management, cognitive radios, interference management, 5G, and Machine Learning.
	\end{IEEEbiography}
	
\begin{IEEEbiography}[{\includegraphics[width=1in,height=1.25in,clip,keepaspectratio]{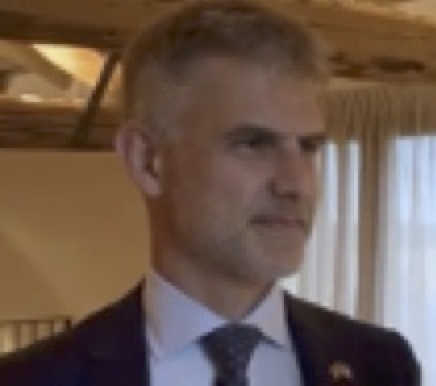}}]
{Alessandro Vanelli-Coralli} (Senior Member, IEEE) received the Dr.Ing. degree in electronics engineer- ing and the Ph.D. degree in electronics and computer science from the University of Bologna, Italy, in 1991 and 1996, respectively. In 1996, he joined the University of Bologna, where he is currently an Associate Professor. He chaired the Ph.D. Board, Electronics, Telecommunications and Information Technologies from 2013 to 2018. From 2003 to 2005, he was a Visiting Scientist with Qualcomm Inc., San Diego, CA, USA. He participates in national and international research projects on wireless and satellite communication systems and he has been a Project Coordinator and scientific responsible for several European Space Agency and European Commission funded projects. He is currently the Responsible for the Vision and Research Strategy task force of the Networld2020 SatCom Working Group. He is a member of the Editorial Board of the Wiley InterScience Journal on Satellite Communications and Networks and Associate Editor of the Editorial Board of Aerial and Space Networks Frontiers in Space Technologies. Dr. Vanelli-Coralli has served in the organisation committees of scientific conferences and since 2010 he is the general CoChairman of the IEEE ASMS Conference. He is co- recipient of several the Best Paper Awards and he is the recipient of the 2019 IEEE Satellite Communications Technical Recognition Award.
	\end{IEEEbiography}

\begin{IEEEbiography}[{\includegraphics[width=1in,height=1.25in,clip,keepaspectratio]{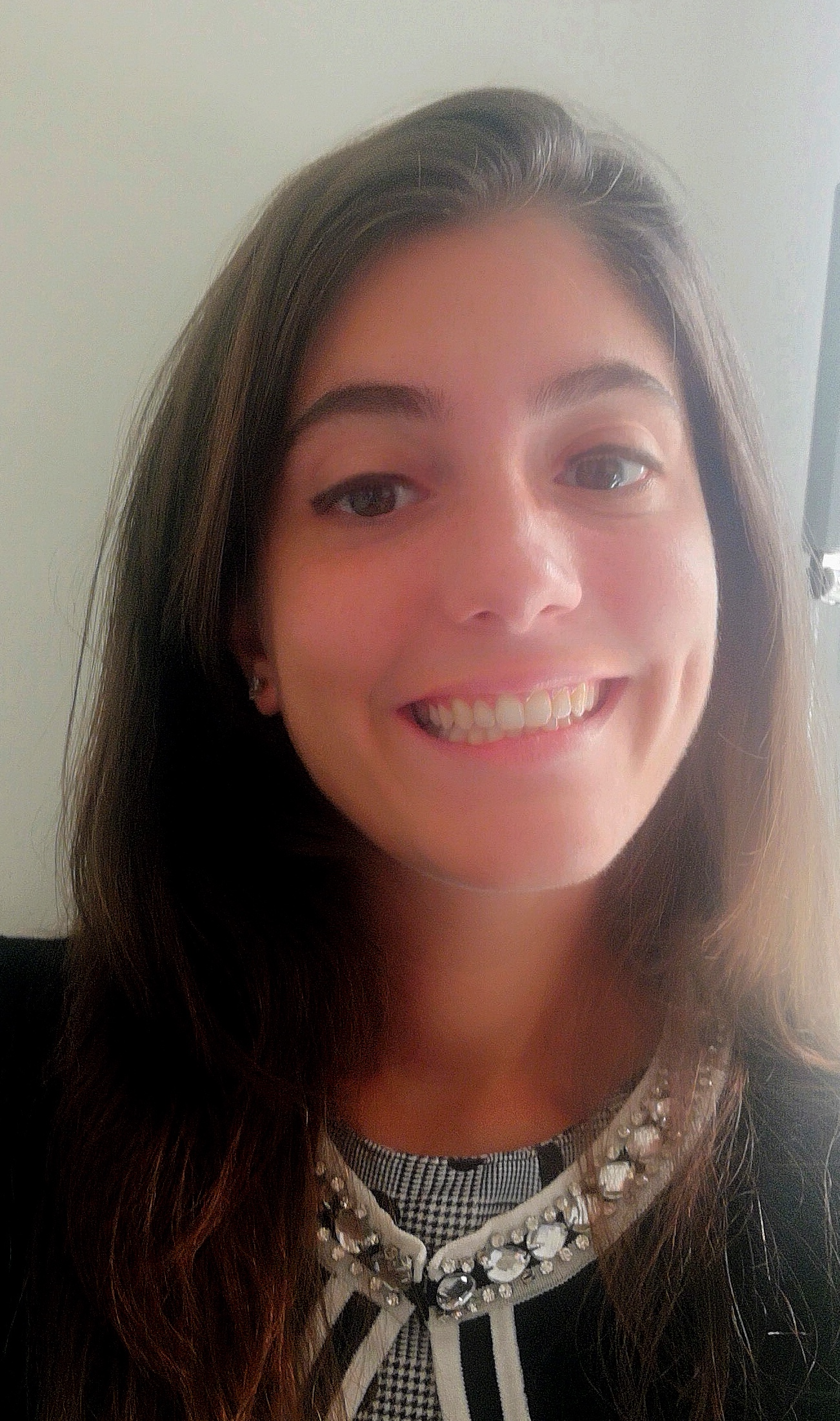}}]
{Carla Amatetti} obtained the Master Degree in Communications and Computer Networks Engineering in 2018 from the Politecnico di Torino, Italy. From January to June 2019, she worked for the Fiat Research Center in Torino, contributing to the application of 5G to vehicular communications, for emergency scenarios. From November 2019, she is purchasing a PhD at the University of Bologna in the Department of Electrical, Electronic and Information Engineering ``Guglielmo Marconi'' (DEI). From January to July 2022 she was a visiting researcher at the Interdisciplinary Centre for Security, Reliability, and Trust (SnT) within the University of Luxembourg.  From July 2019 to November 2019, she received a research grant about ``5G Non-Terrestrial Networks: physical layer design and assessment.'' Her research work is focused on the study of solutions and techniques for the integration of Non Terrestrial and Terrestrial communication systems, with particular emphasis on Narrowband Internet of Things (NB-IoT) technology.
\end{IEEEbiography}

\EOD

\end{document}